\documentclass[11pt,reqno]{amsart}
\usepackage{amsfonts}
\usepackage[framemethod=TikZ]{mdframed}
\usepackage{xcolor}
\usepackage{framed}
\usepackage{geometry}
\usepackage{amsmath,amssymb,amsthm}
\usepackage[authoryear]{natbib}
\usepackage{graphicx}
\usepackage{comment}
\usepackage{tikz}
\usepackage{bm}
\usepackage{epstopdf}
\usepackage{hyperref}
\usepackage{booktabs}
\usepackage{array}

\theoremstyle{plain}
\newtheorem{theorem}{Theorem}

\newtheorem{lemma}{Lemma}
\newtheorem{corollary}{Corollary}

\theoremstyle{definition}
\newtheorem{definition}{Definition}
\newtheorem{remark}{Remark}
\newtheorem{example}{Example}
\newtheorem{algorithm}{Algorithm}

\newcommand{\mK}{\mathcal{K}}
\renewcommand{\qed}{\hfill{\tiny \ensuremath{\blacksquare} }}
\newcommand{\lto}{\leftarrow}

\renewcommand{\Pr}{{\mathrm{P}}}

\numberwithin{equation}{section}
 
\usetikzlibrary {positioning}
\input{tcilatex}

\begin{document}

\title[Uniform Inference on Discrete Quantiles]{Generic Inference on
Quantile and Quantile Effect Functions for Discrete Outcomes}
\thanks{We would like to thank for useful comments and feedback the students
of the courses 14.382 at MIT and EC709 at Boston University, where the ideas
presented here have been taught for several years. We are grateful to the editor, associate editor, five anonymous reviewers, Shuowen Chen and  seminar participants at UC Davis, UC Irvine, UC San Diego, University of Connecticut, and
UCL for helpful comments. We would like to acknowledge the financial support
from the NSF and from the Swiss National Science Foundation for the grant
165621. Data and code for the empirical applications is available at \href{https://github.com/bmelly/discreteQ}{\texttt{https://github.com/bmelly/discreteQ}}.}
\author{Victor Chernozhukov \ Ivan Fernandez-Val \ Blaise Melly \\ Kaspar W%
\"{u}thrich}
\date{\today }

\begin{abstract}

Quantile and quantile effect functions are important tools for descriptive and causal analyses due to their natural and intuitive interpretation. Existing inference methods for these functions do not apply to discrete  random variables. This paper offers a simple, practical
construction of simultaneous confidence bands for quantile and quantile effect functions of possibly discrete random variables.
It is based on a natural transformation of simultaneous confidence bands for distribution functions,
which are readily available for many problems. The construction is generic and does not depend on the nature of the
underlying problem. It works in conjunction with parametric, semiparametric,
and nonparametric modeling methods for observed and counterfactual distributions, and does not depend on the sampling
scheme. We apply our method to characterize the distributional impact of
insurance coverage on health care utilization and obtain the 
distributional decomposition of the racial test score gap.  We find that universal insurance coverage increases the number of doctor visits across the entire distribution, and that the racial test score gap is small at early ages but grows with age due to socio-economic factors affecting child development especially at the top of the distribution. These are new, interesting empirical findings that complement previous analyses that
focused on mean effects only.  In both applications, the outcomes of interest
are discrete rendering existing inference methods invalid  for obtaining
uniform confidence bands for observed and counterfactual quantile functions
and for their difference -- the quantile effects functions.

\medskip

\medskip

\end{abstract}

\maketitle

\section{Introduction}

The quantile function (QF), introduced by \citet{galton1874proposed}, has become a standard tool for descriptive and inferential analysis due to its straightforward and intuitive interpretation. For instance, quantiles play a crucial role in exploratory data analysis as advocated by Tukey and his co-authors. In a classical book, \citet{tukey1977exploratory} promoted the use of the five-number summary, box-plot and  median smoothers, which are all based on quantiles. \citet{doksum1974empirical} and \cite{erich1975nonparametrics} suggested to report the quantile effect (QE) function---the difference between two QFs---to compare the entire distribution of a random variable in two different populations. In randomized control trials and natural experiments, QEs have a causal interpretation, and are usually referred to as quantile treatment effects (QTEs). Quantile regression (QR), introduced by \citet{koenker1978regression}, extends this concept to non-binary treatments and multivariate models. The monograph of \cite{Koenker05} and the handbook edited by \cite{koenker2017handbook} provide  thorough reviews of recent developments of the quantile regression methodology with applications to treatment effects, survival analysis, longitudinal data, time series, and financial data, among others. \citet{chernozhukov+13inference} develop the use of QR and distribution regression (DR) methods to construct counterfactual distribution and quantile functions, which serve as building blocks in policy and decomposition analysis.

In this paper we propose a generic procedure to obtain confidence bands for QFs and QE functions that are valid for continuous, discrete and mixed discrete-continuous outcomes, including counterfactual or latent outcomes. Plotting these bands allows us to  visualize the sampling uncertainty associated with the estimates of these functions. The bands have a straightforward interpretation: they cover the true functions with a pre-specified probability, e.g. 95\%, such that any function that lies outside of the band even at a single quantile can be rejected at the corresponding level, e.g. 5\%. In addition, they are versatile: the same confidence band can be used for testing different null hypotheses. The researcher does not even need to know the hypothesis that will be considered by the reader. For instance, the hypothesis that a treatment has no effect on an outcome can be rejected if the confidence band for the QE function does not cover the zero line. First-order stochastic dominance implies that some non-negative values are covered at all quantiles. The location-shift hypothesis that the QE function is constant implies that there is at least one value covered by the band at all quantiles.

The proposed method relies on a natural transformation of simultaneous confidence bands for distribution functions (DFs) into simultaneous confidence bands for QFs and QE functions. We \textit{invert} confidence bands for DFs (DF-bands) into confidence bands for QFs (QF-bands) and impose shape and support restrictions. We then take the Minkowski difference of the QF-bands, viewed as sets, to construct confidence bands for the QE functions (QE-bands).  This method is generic and  applies to a wide collection of model-based estimators of conditional and marginal  DFs of \textit{discrete}, continuous and mixed continuous-discrete outcomes with and without covariates. The only requirement  is the existence of a valid method for obtaining simultaneous  DF-bands, which is readily available under general sampling conditions for cross-section, time series and panel data. This includes the classical settings of the empirical DFs as a special case, but is much more general than that. For instance, in our empirical applications, we analyze QFs and QE functions that are obtained by inverting counterfactual DFs constructed from regression models with covariates.\footnote{A counterfactual DF is the DF of a potential outcome that is not directly observable but can be constructed using a model. In Section \ref{sec:applications}, the counterfactual DFs are formed by integrating the distribution of an outcome conditional on a vector of covariates in one population with respect to the marginal distribution of the covariates in a different population.} 

Our method can be used to construct three types of bands. First, we show how to invert a DF-band into a QF-band. We prove that there is no loss of coverage by the inversion in that the resulting QF-band covers the entire QF with the same probability as the source DF-band covers the entire DF.  Second, we iterate the method to construct \textit{simultaneous} QF-bands for multiple QFs. Here simultaneity not only means that the bands are uniform---in that they cover the whole function---but also that all the functions are covered by the corresponding bands jointly with the prescribed probability. These bands can be used to test \textit{any} comparison between the QFs such as that differences or ratios of them are constant. By construction, our simultaneous bands are not conservative whenever the source DF-bands are not conservative. 
Third, for the leading case of differences of QFs, we construct QE-bands as differences of QF-bands. 
Our QE-bands can be conservative due to the projection implicit when we take differences of the QF-bands. However, as we discuss below in the literature review, we are not aware of any generic method to construct valid QE-bands for discrete outcomes. We also show how to make the the QF-bands and QE-bands more informative by imposing support restrictions when the outcome is discrete. 
To implement all types of bands, we provide explicit  algorithms based on bootstrap. 

\begin{figure}[tbph]
\begin{center}
\includegraphics[height=7.5in]{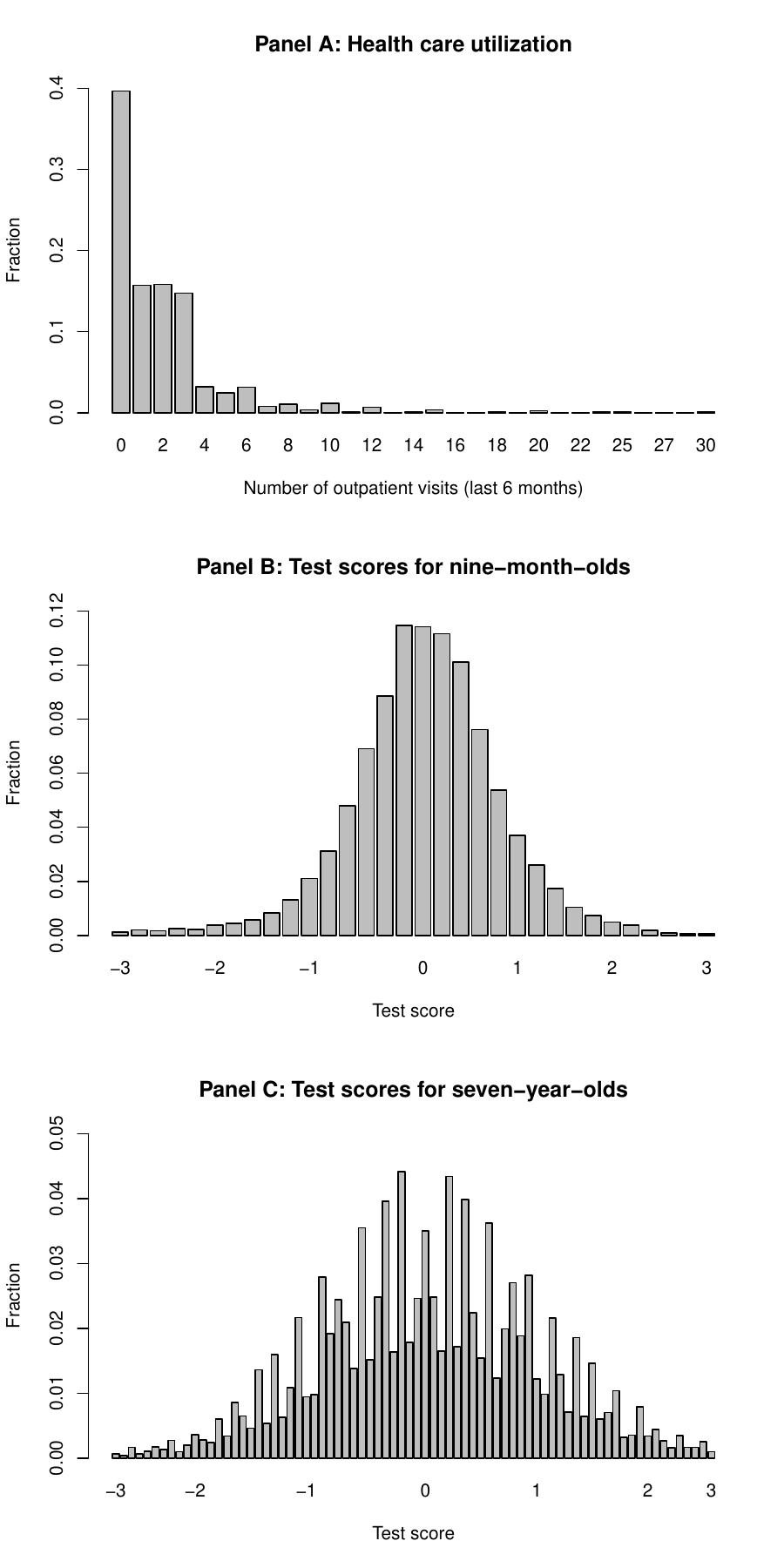}
\end{center}
\caption{Histograms of the outcomes in our empirical examples. Panel A shows
the outcome of our first application reported in Section 4.3; Panel B and C
show the outcomes of our second application reported in Section 4.4. Each
unique value of the variables has been assigned its own bin.}
\label{figure:histograms}
\end{figure}

One important application of our method concerns the estimation of QFs and QE functions from models with covariates, which can be used for causal inference as we show in our empirical examples. When the outcome is continuous, quantile
regression is a convenient model
to incorporate covariates. In many interesting applications, however, the outcome is not continuously distributed. This is naturally the case with count data, ordinal data, and discrete duration data,
but it also concerns test scores that are functions of a finite number of
questions, censored variables, and other mixed discrete-continuous
variables. Examples include the number of doctor visits in our first
application (see Panel A in Figure \ref{figure:histograms}), IQ test scores
for children in our second application (see Panels B and C in Figure \ref%
{figure:histograms}), and wages that have mass points at round values and at
the minimum wage. 
For discrete  outcomes, however, the existing (uniform) inference methods for QR (e.g., \citet{gutenbrunner1992process}, \citet{koenker2002inference}, \citet{angrist2006misspecification}, \citet{qu2015nonparametric}, \citet{belloni2017series}) break down. In addition, the linearity assumption for the conditional quantiles underlying QR is highly implausible in that case. For instance, the linear Poisson regression model does not have linear conditional quantiles.

The  classical models for discrete outcomes such as Poisson, Cox proportional hazard or ordered response models are highly parametrized. These models have the advantage of being parsimonious in terms of parameters and easy to interpret. However, they impose strong homogeneity restrictions on the effects of the covariates. For instance, if a covariate increases the average outcome, then it must increase all the quantiles of the outcome distribution. Moreover, Poisson models imply a restrictive single crossing property on the sign of the estimated probability effects \citep{winkelmann06}. To avoid these limitations we employ the DR model in our applications. \citet{williams1972analysis} introduced this model to analyze ordered outcomes without assuming proportional odds. They estimated a different binary logistic regression for each category of the outcome instead of assuming that the slope coefficients are the same for all categories. \citet{foresiperacchi95} estimated a sequence of logistic regression to obtain the conditional distribution of excess  returns at a fixed number of thresholds.    \citet{chernozhukov+13inference}  considered a continuum of binary regressions and showed that the continuum provides a coherent
and flexible model for the entire conditional distribution. \citet{rothe2013testing} proposed specification tests for DR. 
 DR is a comprehensive tool for modeling and estimating the entire
conditional distribution of any type of outcome (discrete, continuous, or
mixed). It allows the covariates to affect differently the outcome at
different points of the distribution and encompasses several  classical parametric models as special cases (classical linear regression, Cox proportional hazard, Poisson regression).  

The cost of this flexibility is that
the DR parameters can be hard to interpret because they do not correspond to
QEs. To overcome this problem, we propose to report QEs computed as
differences between the QFs of counterfactual distributions estimated by DR
in conjunction with confidence bands constructed using our projection
method. These one-dimensional functions provide an intuitive summary of the
effects of the covariates. We argue that this combination of our generic procedure with the DR model
provides a comprehensive and practical approach for estimating QFs and QE
functions with discrete data that can be utilized for causal inference. While we focus on DR in this paper, we emphasize that our method
also combines well with classical parametric models such as Poisson, Cox proportional hazard and ordered response
regression models.
To the best of out knowledge, there is no inference method available to construct valid QF-bands and QE-bands even in these simple models. 
In addition, our method
works in conjunction with more recent inference approaches for DFs with
potentially discrete data. Examples include \citet{frandsen2012rdd}, %
\citet{donald2014estimation}, \citet{hsu2015estimation}, and %
\citet{belloni2017program}. 

We apply our approach to two problems, featuring two common types of
discrete outcomes. In the first application, we exploit a large-scale
randomized control trial in Oregon to estimate the distributional impact of
universal insurance coverage on health care utilization measured by the number of
doctor visits. Since this outcome is a count, we estimate the conditional
DFs using both Poisson and distribution regressions. Poisson regression
clearly underestimates the probability of having zero visits as well as that
of having a large number of visits. The more flexible DR finds a positive
effect, especially at the upper tail of the distribution. This is an
interesting empirical finding in its own right; it complements the mean
regression analysis results reported in \citet{finkelstein+12}. In the
second application, we reanalyze the racial test score gap of young
children. As shown in Figure \ref{figure:histograms}, test scores are discrete.  We find that while there is very little gap at eight months, a
large gap arises at seven years. In addition, looking at the whole
distribution, we uncover that the observed racial gap is widening in the
upper tail of the distribution of test scores. The increase in the gap can
be mostly explained by differences in socio-economic factors affecting the development of the child as captured by observed covariates. These
results complement and expand the findings of \citet{fryer2013testing} for
the mean racial test score gap, revealing what happens to the entire
distribution.

\paragraph{\textbf{Literature Review}} 
To the best of our knowledge, this is the first paper that provides asymptotically similar simultaneous QF-bands for discrete outcomes. \citet{scheffe1945non} were the first to consider
empirical quantiles for discrete data. They showed that pointwise confidence
intervals obtained by inverting pointwise confidence intervals for
the DF based on the empirical DF are still valid but conservative when the outcome is discrete.\footnote{\citet{scheffe1945non} studied the properties of confidence intervals for quantiles based on order statistics. This way of obtaining confidence intervals is equivalent to inverting confidence intervals for the DF. \cite{woodruff1952confidence} and \cite{francisco1991quantile} graphically illustrated and formally defined, respectively, the inversion idea. However, their formal results only apply to continuous outcomes.} 
\citet{frydman2008discrete} and \citet{larocque2008confidence} suggested
methods to obtain the exact coverage rate of these confidence intervals.
In contrast, our confidence bands for the QFs are uniform in the probability index and not conservative, and can be based on more general estimators of the DF.

Another strand of the
literature tried to overcome the discreteness in the data by adding a small
random noise to the outcome (also called jittering), see for instance %
\citet{machado2005quantiles} and the applications in %
\citet{koenker2002inference} and \citet{chernozhukov+13inference}. %
\citet{ma+11asymptotic} considered an alternative definition of quantiles
based on linearly interpolated DFs. These strategies restore asymptotic
Gaussianity of the empirical QFs and QE functions, at the price of changing
the estimand. One might argue that this change is not a serious issue when
the number of points in the support of the outcome is large, but we find it
more transparent to work directly with the observed discrete outcome. Thus, we keep the focus on the original QE function at the price that our QE-bands might be conservative. However, we find that they lead to informative inferences in two empirical applications and in extensive numerical simulations, where they are not conservative in most cases.  Since we are not aware of any alternative generic method to construct asymptotically similar QE-bands of discrete outcomes, we believe this is a useful addition to the statistical toolkit. 

Our method can also be applied to obtain QF-bands and QE-bands of continuous outcomes, complementing the existing well-established methods  for this type of outcomes. For example, \citet{kh15} provided a method to construct exact
DF-bands and QF-bands based on the empirical distribution of independent and identically distributed data.\footnote{We refer to \citet{kh15} for an excellent review on construction of confidence bands from the empirical distribution function.}  Our method has the more modest goal of constructing asymptotic confidence bands, but applies more generally to any estimator of the distribution that obeys a functional central limit theorem. This includes Poisson, Cox proportional hazard and DR-based estimators of the distribution from weakly dependent data. \citet{chernozhukov+13inference} also used DR as the basis for constructing QF-bands and QE-bands of continuous outcomes. Their construction consists of two steps. First, obtain estimators of the QFs and QE functions from estimators of DFs by inversion. Second, construct QF-bands and QE-bands from the limit distributions of the estimators of the QFs and QE functions derived from the limit distribution of the estimators of the distribution functions via delta method. This construction has the advantage of producing asymptotically similar QE-bands, but  breaks down for discrete outcomes because the quantile (left-inverse) mapping is not smooth (Hadamard differentiable), which precludes the application of the delta method in the second step. Our construction of the confidence bands consists of similar steps but the order of the steps is different. First, we construct DF-bands using the limit distribution of the estimators of the DFs. Second, we construct the QF-bands by
inversion  the DF-bands and take differences to construct the QE-bands. This difference in the order of the steps completely avoids the delta method and is the key to apply our method to discrete outcomes. Hence, we see our method as complementary to \citet{chernozhukov+13inference}. 

Finally, we would like to emphasize that none of the existing methods that we are aware of can be applied to construct valid QF-bands and QE-bands in our two empirical applications where the outcomes are discrete and  the estimators of the distribution are model-based.

\paragraph{\textbf{Outline}} The rest of the paper is organized as follows. Section \ref{sec:bands}
introduces our generic method to construct 
QF-bands and QE-bands. Section \ref{section:algorithms} provides an explicit
algorithm based on bootstrappable estimators for the DFs. Section \ref%
{sec:applications} presents the two empirical applications. Appendix \ref%
{app:CBDF} shows how to improve the finite sample properties of DF-bands by imposing logical monotonicity or range restrictions.
Appendix \ref{app:algorithms_single_qf} provides an additional algorithm to
construct confidence bands for single QFs. Appendix \ref{app:sim}  reports the results of a simulation study.

\section{Generic Confidence Bands for Quantile and Quantile Effect Functions}

\label{sec:bands}

This section contains the main theoretical results of the paper. Our only
assumption is the availability of simultaneous confidence bands for DFs.
Since the seminal work of \citet{kolmogoroff1933}, various methods to obtain simultaneous confidence bands have been developed.\footnote{The original Kolmogorov bands are actually conservative for discrete random
variables, see \citet{kolmogoroff1941confidence}. Alternative methods, such
as those described in Section \ref{section:algorithms}, are asymptotically
exact.} In Section \ref{section:algorithms} we provide an algorithm to construct simultaneous DF-bands
that can be applied when the estimators of the DFs are known to be
bootstrappable, which is often the case.


\subsection{Confidence Bands for Distribution Functions}

\label{subsec:CBDF}

Let $\mathcal{Y}$ be a closed subinterval in the extended real number line $%
\overline{\mathbb{R}} = \mathbb{R} \cup \{-\infty,+\infty\}$. 
Let $\mathbb{D}$ denote the set of nondecreasing functions, mapping $%
\mathcal{Y}$ to $[0,1]$. A function $F$ is
nondecreasing if for all $x,y \in \mathcal{Y}$ such that $x \leq y$, one has $%
F(x) \leq F(y)$. We will call the elements of the set $\mathbb{D}$  \textquotedblleft
distribution functions", albeit some of them need not be proper DFs. In what follows, we let $%
F$ denote some target DF. $F$ could be a
conditional DF, a marginal DF, or a counterfactual DF. 

\begin{definition}[DF-Band of Level $p$]
Given two functions $y \mapsto U(y) $ and $y \mapsto L(y) $ in the set $%
\mathbb{D}$ such that $L \leq U$, pointwise, we define a band $I = [L,U]$ as
the collection of intervals 
\begin{equation*}
I(y) = [L(y), U(y)], \quad y \in \mathcal{Y}.
\end{equation*}
We say that $I$ covers $F$ if $F \in I$ pointwise, namely $F(y) \in I(y)$
for all $y \in \mathcal{Y}$. If $U$ and $L$ are some data-dependent bands,
we say that $I=[L,U]$ is a DF-band of level $p$, if $I$
covers $F$ with probability at least $p$. \qed
\end{definition}

In many applications the point estimates $\hat{F}$ and confidence bands $%
[L^{\prime },U^{\prime }]$ for the target distribution $F$ do not satisfy
logical monotonicity or range restrictions, namely they do not take values
in the set $\mathbb{D}$. Appendix \ref{app:CBDF} shows that given such an
ordered triple $L^{\prime }\leq \hat{F}\leq U^{\prime }$, we can always
transform it into another ordered triple $L\leq \check{F}\leq U$ that obeys
the logical monotonicity and range restrictions. Such a transformation will
generally improve the finite sample properties of the point estimates and
confidence bands.

\subsection{Confidence Bands for a Single Quantile Function}

\label{subsec:CBQF} Here we discuss the construction of confidence bands for
the left-inverse function of $F$, $F^{\leftarrow}$, which we call ``quantile
function'' of $F$.

\begin{definition}[Left-Inverse]\label{def:linv}
Given a function $y \mapsto G(y)$ in $\mathbb{D}$, we define its left-inverse by $G^{\leftarrow }(a):=\inf \{y\in \mathcal{Y}:G(y)\geq a\}
$ if  $\sup_{y \in \mathcal{Y}} G(y) \geq a$  and $G^{\leftarrow }(a): = 
\sup \{ y \in \mathcal{Y} \} $ otherwise.
\end{definition}

The following theorem provides a confidence band $I^{\leftarrow
}$ for the QF $F^{\leftarrow }$ based on a generic confidence band $I$ for $%
F $. 

\bigskip

\begin{theorem}[Generic QR-Band]
\label{theorem:bquant} Consider a distribution function $F$ and band
functions $L$ and $U$ in the class $\mathbb{D}$. Suppose that the
distribution function $F$ is covered by $I$ with probability $p$. Then, the quantile function $F^{\leftarrow }$ is covered by $I^{\leftarrow }$
with probability $p$, where 
\begin{equation*}
I^\leftarrow(a): = [U^\leftarrow(a), L^\leftarrow(a)], \ \ a \in [0,1].
\end{equation*}

%
\end{theorem}

\textbf{Proof.} Here we adopt the convention $\inf\{\emptyset\} = + \infty$ so that the left-inverse function can be defined as $G^{\leftarrow }(a):=\inf \{y\in \mathcal{Y}:G(y)\geq a\}\wedge
\sup \{y\in \mathcal{Y}\}$, which avoids distinguishing the two cases of Definition \ref{def:linv}. 

It suffices to show that $U^\leftarrow \le F^\leftarrow \le L^\leftarrow$ if and only if $L \le F \le U$.
We first show that $F^\leftarrow \le L^\leftarrow$ if and only if $F \ge L$. For the ``if'' part, note that for any $a\in \lbrack 0,1]$, since $L(y)\leq F(y)$ for each $y\in \mathcal{Y}
$ and $F,L \in \mathbb{D}$, 
\begin{eqnarray*}
F^{\leftarrow }(a) &=&\inf \{y\in \mathcal{Y}:F(y)\geq a\}\wedge \sup 
\mathcal{Y} \\
&\leq &\inf \{y\in \mathcal{Y}:L(y)\geq a\}\wedge \sup \mathcal{Y}%
=L^{\leftarrow }(a).
\end{eqnarray*}%
For the ``only if'' part, we use that for any $G \in \mathbb{D}$, $G^{\to} \circ G^{\lto} = G$, where $G^{\to}$ denotes the right-inverse of $G^{\lto}$ defined by
$$
G^{\to} \circ G^{\lto} (y) := \sup\{a \in [0,1] : G^{\lto}(a) \leq y \} \vee 0, 
$$
where we use the convention $\sup\{\emptyset\} = - \infty$. Then,  for any $y\in \mathcal{Y}$, since $F^{\leftarrow}(a)\le L^{\leftarrow}(a)$ for each $a\in [0,1]$ and $a \mapsto F^{\lto}(a)$ and $a \mapsto L^{\lto}(a)$ are nondecreasing, 
\begin{eqnarray*}
F(y) = F^{\to} \circ F^{\lto} (y) &=&\sup \{a\in [0,1]:F^{\leftarrow}(a)\le y\}\vee 0 \\
&\ge &\sup \{a\in [0,1]:L^{\leftarrow}(a)\le y\}\vee 0 = L^{\to} \circ L^{\lto}(y)  = L(y).
\end{eqnarray*}%
Analogously, we can conclude that $F^\leftarrow \ge U^\leftarrow$ if and only if $F \le U$. \qed

%
%

\begin{remark}[Similarity]
 The QF-band $I^{\leftarrow }$ is constructed by applying the left-inverse transformation to the
DF-band $I$.  Theorem \ref{theorem:bquant} shows that there is no loss of coverage in the inversion of the band. Hence, our generic method carries over the similarity (non-conservativeness) of the
DF-band to the QF-band.  
\end{remark}

We can narrow $I^{\leftarrow }$ without affecting its coverage by
exploiting the support restriction that the quantiles can only take the
values of the underlying random variable. This is relevant when the variable
of interest is discrete as in the applications presented in Section \ref%
{sec:applications}. Suppose that $T$ is the support of the random variable
with DF $F$. 
Then it makes sense to exploit the support restriction that $F^{\leftarrow
}(a)\in T$ by intersecting the confidence band for $F^{\leftarrow }$ with $%
T $. Clearly, this will not affect the coverage properties of the bands.

\begin{corollary}[Imposing Support Restrictions]
\label{corollary:support} Consider the set $\tilde I^\leftarrow$ defined by
pointwise intersection of $I^\leftarrow$ with $T$, namely $\tilde
I^\leftarrow(a):= I^\leftarrow(a) \cap T.$ Then, $\tilde I^\leftarrow
\subseteq I^\leftarrow$ pointwise, and if $I^\leftarrow$ covers $%
F^\leftarrow $ then so does $\tilde I^\leftarrow$.
\end{corollary}

\begin{figure}[tbph]
\begin{center}
\includegraphics[height=2.2in]{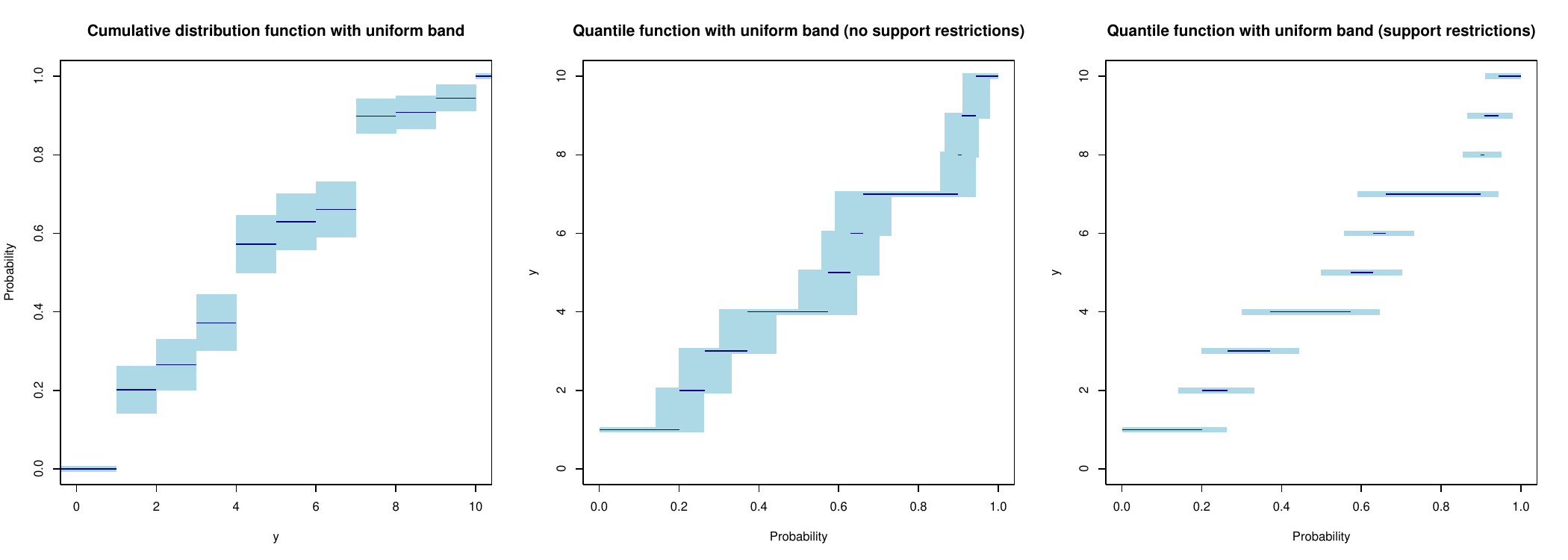}
\end{center}
\caption{{Construction of QF-bands using Theorem \protect\ref%
{theorem:bquant} and Corollary \protect\ref{corollary:support}. Left: DF $F$ (dark line) and DF-band $I$ (light
rectangles). Middle: QF $F^{\leftarrow }$ and QF-band $I^{\leftarrow }$. Right: The support-restricted QF-band $\tilde{I}^{\leftarrow }$}}
\label{figure:illustration}
\end{figure}

Figure \ref{figure:illustration} illustrates the construction of bands using
Theorem \ref{theorem:bquant} and Corollary \ref{corollary:support}. The left
panel shows a DF $F:[0,10]\mapsto \lbrack 0,1]$ covered by a DF-band $I=[L,U]$.
The middle panel shows that the inverse map $F^{\leftarrow }:[0,1]\mapsto
\lbrack 0,10]$ is covered by the inverted band $I^{\leftarrow
}=[U^{\leftarrow },L^{\leftarrow }]$. The band $I^{\leftarrow }$ is easy to
obtain by rotating and flipping $I$, but does not exploit the fact that the
support of the variable with distribution $F$ in this example is the set $%
T=\{0,1,\ldots ,10\}$. By intersecting $I^{\leftarrow }$ with $T$ we obtain
in the right panel the QF-band $\tilde{I}^{\leftarrow }$ which reflects the
support restrictions.

\subsection{Generic Confidence Bands for Multiple Quantile Functions and Quantile Effects}

\label{subsec:CBQE}

The quantile effect (QE) function $a\mapsto \Delta _{j,m}(a)$ is the
difference between the QFs of two random variables with DFs $F_{j}$ and $%
F_{m}$ and support sets $T_{j}$ and $T_{m}$, i.e., 
\begin{equation*}
\Delta _{j,m}(a):=F_{j}^{\leftarrow }(a)-F_{m}^{\leftarrow }(a),\ \ a\in
\lbrack 0,1].
\end{equation*}%
Our next goal is to construct simultaneous confidence bands that jointly
cover the DFs, $\left( F_{k}\right) _{k\in \mathcal{K}}$, 
the corresponding QFs, $\left( F_{k}^{\leftarrow}\right) _{k\in \mathcal{K}}$, and the QE functions, $\left( \Delta _{j,m}\right)
_{\left( j,m\right) \in \mathcal{K}^{2}}$, where $\mathcal{K}$ is a finite
set. For example, $\mathcal{K}=\{0,1\}$ (treated and control outcome
distributions) in our first application and $\mathcal{K}=\{W,B,C\}$ (white,
black and counterfactual test score distributions) in our second application.

Specifically, suppose we have the DF-bands $\left( I_{k}\right)
_{k\in \mathcal{K}}$, which jointly cover the DFs $\left( F_{k}\right)
_{k\in \mathcal{K}}$ with probability at least $p$. For example, we can
construct these bands using Theorem \ref{theorem:bquant} in conjunction with
the Bonferroni inequality.\footnote{The joint coverage of two confidence bands with marginal coverage
probabilities $\tilde{p}$ is at least $p=2\tilde{p}-1$ by Bonferroni
inequality.} Alternatively, the generic Algorithm \ref{algorithm:qte} presented in Section %
\ref{section:algorithms} provides a construction of a joint confidence band
that is not conservative. First we construct the QF-bands $\left(
I_{k}^{\leftarrow}\right) _{k\in \mathcal{K}}$, which jointly cover the QFs $%
\left( F_{k}^{\leftarrow}\right) _{k\in \mathcal{K}}$ with probability at
least $p$ by Theorem \ref{theorem:bquant}. Then we convert these bands to
confidence bands for $\left(\Delta _{j,m}\right)_{(j,k)\in \mathcal{K}^{2}}$
by taking the pointwise Minkowski difference $\ominus$ of each of the pairs
of the two bands, viewed as sets. Recall that the Minkowski difference
between two subsets $V$ and $U$ of a vector space is $V\ominus U:=\{v-u:v\in
V,u\in U\}$. We note that if $V$ and $U$ are intervals, $[v_{1},v_{2}]$ and $%
[u_{1},u_{2}]$, then 
\begin{equation*}
V\ominus U=[v_{1},v_{2}]\ominus \lbrack
u_{1},u_{2}]=[v_{1}-u_{2},v_{2}-u_{1}].
\end{equation*}%
In words, the upper-end of the interval for the difference $v-u$ is the difference between the upper-end of the interval for $v$ and the lower-end of the interval for $u$. Symmetrically, the lower-end of the interval for the difference $v-u$ is the difference between the lower-end of the interval for $v$ and the upper-end of the interval for $u$. This greatly simplifies the practical computation of the bands.

\begin{theorem}[Generic Simultaneous QF-Bands and QE-Bands]
\label{theorem:bqte} Consider the distribution functions $\left(F_{k}\right)
_{k\in \mathcal{K}}$ and the band functions $\left(I_{k}:=[L_{k},U_{k}]%
\right) _{k\in \mathcal{K}}$ in the class $\mathbb{D}$. Suppose that the
distribution functions $\left( F_{k}\right) _{k\in \mathcal{K}}$ are jointly
covered by $\left( I_{k}\right) _{k\in \mathcal{K}}$ with probability $p$.
Then:

\begin{enumerate}
\item The quantile functions $\left( F_{k}^{\lto}\right) _{k\in \mathcal{K}}$ are jointly
covered by $\left( I_{k}^{\lto}\right) _{k\in \mathcal{K}}$ with probability $p$, where 
$I_{k}^{\lto} = [U_{k}^{\leftarrow},L_{k}^{\leftarrow }]$.
\item For $(j,k)\in \mathcal{K}^2$, the quantile effect function, $\Delta
_{j,m}$, is covered by $I_{\Delta \left( j,m\right) }^{\leftarrow
}=[U_{j}^{\leftarrow},L_{j}^{\leftarrow }]-[U_{m}^{\leftarrow
},L_{m}^{\leftarrow }]$ with probability at least $p$, where the minus
operator is defined by a pointwise Minkowski difference: 
\begin{equation*}
I_{\Delta \left( j,m\right) }^{\leftarrow }(a):=[U_{j}^{\leftarrow
}(a),L_{j}^{\leftarrow }(a)]\ominus \lbrack U_{m}^{\leftarrow
}(a),L_{m}^{\leftarrow }(a)],\ \ a\in \lbrack 0,1].
\end{equation*}

\item The confidence bands have the following joint coverage property: 
\begin{equation*}
\Pr (F_{k}\in I_k,F_{k}^{\leftarrow }\in I_{k}^{\leftarrow },\Delta
_{j,m}\in I_{\Delta \left( j,m\right) }^{\leftarrow };\text{ for all }%
(k,j,m)\in \mathcal{K}^{3})=p.
\end{equation*}
\end{enumerate}


\end{theorem}

\textbf{Proof.} The results follows from the definition of the Minkowski
difference and because the event $\cap _{k\in \mathcal{K}}\{F_{k}\in I_k\}$
is equivalent to the event $\cap _{k\in \mathcal{K}}\{F_{k}^{\leftarrow }\in
I_{k}^{\leftarrow }\}$ by Theorem \ref{theorem:bquant}, which implies the
event $\cap _{\left( j,m\right) \in \mathcal{K}^{2}}\{F_{j}^{\leftarrow
}-F_{m}^{\leftarrow }\in I_{\Delta \left( j,m\right) }^{\leftarrow }\}$. 
\qed

Theorem \ref{theorem:bqte} shows that simultaneous QF-bands can be obtained by inverting simultaneous DF-bands, and  QE-bands by taking the Minkowski difference between the two simulataneous 
QF-bands for the corresponding QFs. As in Theorem \ref%
{theorem:bquant}, we can narrow the band $I_{\Delta }^{\leftarrow }$ without
affecting coverage by imposing support restrictions as demonstrated in
Corollary \ref{corollary:support_qte}. 

\begin{corollary}[Imposing Support Restrictions]
\label{corollary:support_qte} For $\left( j,m\right) \in \mathcal{K}^{2}$,
consider the bands $\tilde{I}_{\Delta \left( j,m\right) }^{\leftarrow }=%
\tilde{I}_{j}^{\leftarrow }-\tilde{I}_{m}^{\leftarrow }$ defined by: 
\begin{equation*}
\tilde{I}_{\Delta \left( j,m\right) }^{\leftarrow }(a):=\tilde{I}%
_{j}^{\leftarrow }(a)\ominus \tilde{I}_{m}^{\leftarrow }(a),\quad \tilde{I}%
_{k}^{\leftarrow }(a):=\{[U_{k}^{\leftarrow }(a),L_{k}^{\leftarrow }(a)]\cap
T_{k}\},\quad k\in \mathcal{K}.
\end{equation*}%
Then $\tilde{I}_{\Delta \left( j,m\right) }^{\leftarrow }\subseteq I_{\Delta
\left( j,m\right) }^{\leftarrow }$, and if $I_{\Delta \left( j,m\right)
}^{\leftarrow }$ covers $\Delta _{j,m}$ then so does $\tilde{I}_{\Delta
\left( j,m\right) }^{\leftarrow }$.
\end{corollary}

\medskip

\begin{remark}[Joint Support Restrictions]
The band $\tilde{I}_{\Delta \left( j,m\right) }^{\leftarrow }$ can be
further narrowed if the two random variables with distributions $F_{j}$ and $%
F_{m}$ have restrictions in their joint support $T_{jm}$, i.e., $T_{jm}\neq
T_{j}\times T_{m} =\{(t_{j},t_{m}):t_{j}\in T_{j},t_{m}\in T_{m}\}$. In this
case we can drop all the elements $d$ from $\tilde{I}_{\Delta \left(
j,m\right) }^{\leftarrow }$ that cannot be formed as $d=t_{j}-t_{m}$ for
some $(t_{j},t_{m})\in T_{jm}$. For example, let $T_{j}=T_{m}=\tilde{I}%
_{\Delta \left( j,m\right) }^{\leftarrow }=\{0,1,2\}$, then we can drop $%
\{2\}$ from $\tilde{I}_{\Delta \left( j,m\right) }^{\leftarrow }$ if $%
(t_j,t_m) = (2,0)\notin T_{jm}$. 
\qed
\end{remark}

\medskip

\begin{remark}[Similarity]
Part 1 of Theorem \ref{theorem:bqte} provides a construction of simultaneous QF-bands. These bands can be used to test any comparison between two or more QFs. These include that the difference between each pair of functions is zero or constant, or that all the ratios between each pair of functions is one or constant (see Remark \ref{rem:ratios}). 

Part 3 shows that our generic method of
constructing bands carries over the similarity (non-conservativeness) of the
DF-bands to the simultaneous QF-bands and QE-bands.
Moreover, our construction is optimal in the sense that if we want to
simultaneously cover all the DFs, QFs and QE functions of
interest, it is not possible to construct uniformly shorter bands while
preserving the joint coverage rate once all the joint support restrictions
are imposed. It is common to report at the same time several QFs and QE functions.
For instance, Figures \ref{figure:results_testscores_9m} and \ref%
{figure:results_testscores_7y} provide three different QFs (two observed and
one counterfactual) and the differences between these functions, which are
all of interest. Theorem \ref{theorem:bqte} (together with Corollary \ref%
{cor3} for the asymptotic similarity of the bands for the DFs) shows that
our bands jointly cover asymptotically all these functions with probability $%
p$. This allows for a transparent and honest assessment of hypotheses about
these functions.

On the other hand, when the goal is to cover only a single QE function
independently from the other functions, then our QE-band can
be marginally conservative (Part 2 of Theorem \ref{theorem:bqte}). This is due to the
projection implicit in the application of the Minkowski difference and is
the price to pay for the joint uniform coverage property. However, our
empirical results in Section \ref{sec:applications} and numerical simulations in Appendix \ref{app:sim} clearly demonstrate the usefulness of these bands that
allow for testing hypotheses that could not be considered using existing
methods.  We are not aware of any generic method to construct nonconservative
 QE-bands of discrete outcomes. \qed
 \end{remark}

\begin{figure}[tbph]
\begin{center}
\includegraphics[height=2.2in]{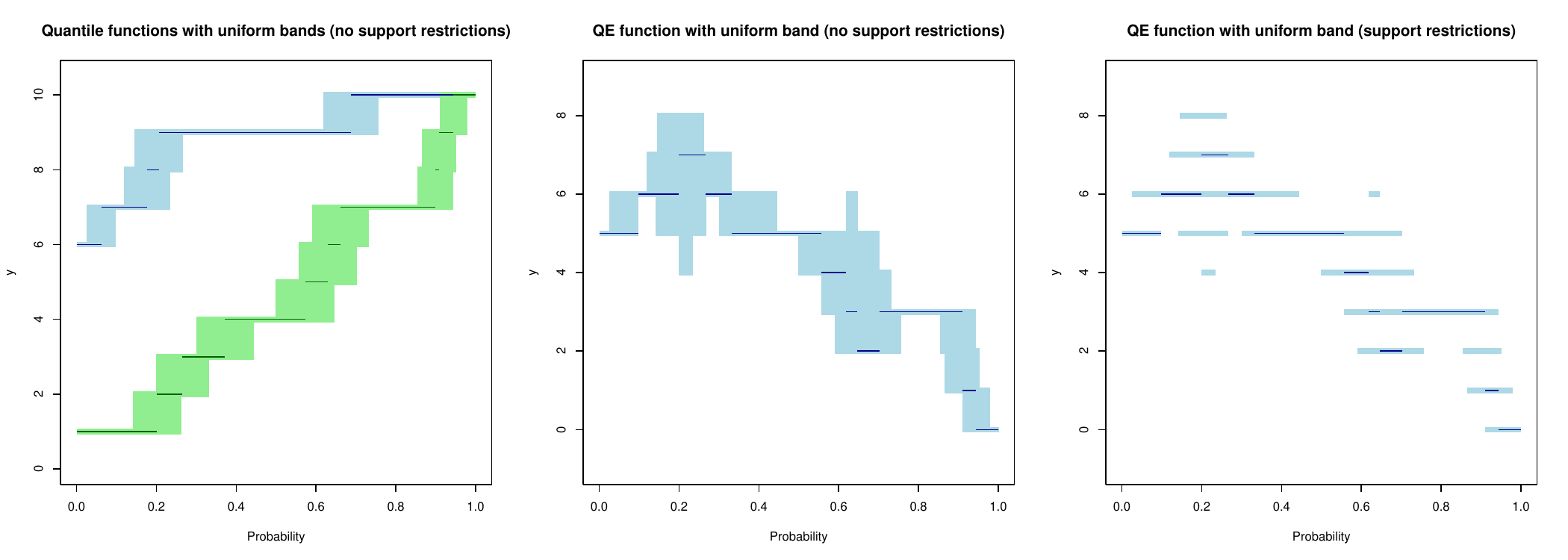}
\end{center}
\caption{{Construction of the QE-bands using Theorem \protect
\ref{theorem:bqte} and Corollary \protect\ref{corollary:support_qte}. Left:
QFs $F^{\leftarrow}_0$ and $F^{\leftarrow}_1$ and QF-bands $I^{\leftarrow}_0$ and $I^{\leftarrow}_1$. Middle: the QE function $%
\Delta$ and the QE-band $\bar{I}_\Delta$ without support
restrictions. Right: the QE function $\Delta$ and the QE-band $\bar{I%
}_\Delta$} with support restrictions.}
\label{figure2}
\end{figure}

Figure \ref{figure2} illustrates the construction of QE-bands using Theorem \ref{theorem:bqte} and Corollary \ref%
{corollary:support_qte}. The left panel shows the bands $I_{0}^{\leftarrow }$
and $I_{1}^{\leftarrow }$ for the QFs $F_{0}^{\leftarrow }$ and $%
F_{1}^{\leftarrow }$. The middle panel shows the band $I_{\Delta \left(
1,0\right) }$ for the QE function $\Delta _{1,0}=F_{1}^{\leftarrow
}-F_{0}^{\leftarrow }$, obtained by taking the Minkowski difference of $%
I_{1}^{\leftarrow }$ and $I_{0}^{\leftarrow }$. The right panel shows the
confidence band $\tilde{I}_{\Delta \left( 1,0\right) }$ for the QE function $%
\Delta _{1,0}$ resulting from imposing the support restrictions. As the
Theorem \ref{theorem:bqte} proves, the QE function $\Delta _{1,0}$ is
covered by the QE-band $I_{\Delta \left( 1,0\right) }$.

\begin{remark}[Confidence Bands for Ratios of QFs]\label{rem:ratios} Theorem \ref{theorem:bqte} provides an explicit construction of bands for differences of QFs, the leading example of comparisons between QFs. Similar simple bands can be constructed for other comparisons of QFs. For example, a confidence band for the ratio of QFs, $\rho_{j,m}(a):= F_{j}^{\leftarrow
}(a)/F_{m}^{\leftarrow }(a)$, can be formed as $I_{\rho \left( j,m\right) }^{\leftarrow
}=[U_{j}^{\leftarrow},L_{j}^{\leftarrow }]/[U_{m}^{\leftarrow},L_{m}^{\leftarrow }]$, where the division
operator is defined pointwise by: 
\begin{equation*}
I_{\rho \left( j,m\right)}(a) := [U_{j}^{\leftarrow}(a)/L_{m}^{\leftarrow }(a), L_{j}^{\leftarrow }(a)/U_{m}^{\leftarrow
}(a)],\ \ a\in \lbrack 0,1].
\end{equation*}
\end{remark}

\section{Computation of Simultaneous Confidence Bands for Distribution Functions}

\label{section:algorithms}In Section \ref{sec:bands} we assumed the
existence of simultaneous DF-bands. Here we describe an
 algorithm that is shown to provide asymptotically valid simultaneous
bands for any bootstrappable estimator of the DFs. Many commonly used
estimators of the DF are bootstrappable under suitable conditions. For
example, \citet{chernozhukov+13inference} give
conditions for bootstrap consistency for the DR-based estimators that we use
in the empirical applications. Maximum likelihood estimators, such as the
Poisson regression that we use as a benchmark in the first application, are
also bootstrappable under weak differentiability conditions, see %
\citet{arcones1992bootstrap}. We note that if the data are discrete, these
existing results still yield validity of bootstrapping  the DF estimator to construct DF-bands, but they do \emph{not} justify the validity of  bootstrapping the QF and QE estimators to construct QF-bands and QE-bands. The reason is that the delta-method breaks down because the left-inverse mapping is no longer (Hadamard) differentiable.

Algorithm \ref{algorithm:qte} provides simultaneous confidence bands that
asymptotically jointly cover the DFs $\left( F_{k}\right) _{k\in \mathcal{K}%
} $, the corresponding QFs $\left( F^{\leftarrow}_{k}\right) _{k\in \mathcal{%
K}}$, and the QE functions $F_{j}^{\leftarrow }-F_{k}^{\leftarrow }$ for all 
$(j,k)\in \mathcal{K}^{2},$ with probability $p$. 
In practice, we estimate the DFs on a grid of points. Let $T$ be a finite
subset of $\mathcal{Y}$. For the DF of a discrete random variable $Y$ with
finite support, we can choose $T$ as the support of $Y$. Otherwise, we can
set $T$ as a grid of values covering the region of interest of the support of $Y$.

\begin{framed}
\begin{algorithm}[Bootstrap Algorithm for QF-bands and QE-bands]  \text{ } 
\label{algorithm:qte}
\begin{enumerate} 
\item Obtain many bootstrap draws  of the estimator   $(\hat{F}_k)_{k \in \mK}$,
$$(\hat{F}^{*(j)}_k)_{k \in \mK}, \quad  j=1, \ldots, B,$$
 where the index $j$ enumerates
the bootstrap draws and $B$ is the number of bootstrap draws (e.g., $B = 1,000$). 

\item For each $y \in T$ and $k \in \mK$, 
compute the  robust standard error of $\hat{F}_k(y)$:
$$
\hat s_k(y) =  (\hat Q_k(.75,y)- \hat Q_k(.25,y))/(\Phi^{-1}(.75) - \Phi^{-1}(.25)),$$
where  $\hat Q_k(\alpha,y)$ denotes the empirical $\alpha$-quantile of the bootstrap sample $(\hat{F}_k^{*(j)}(y))_{j=1}^B$, and $\Phi^{-1}$ denotes the inverse of the standard normal distribution.

\item Compute the critical value 
$$c(p) = \text{ $p$-quantile of  } \left \{ \max_{y \in T, k \in \mK} |\hat{F}_k(y)^{*(j)} - \hat{F}_k(y)|/\hat s_k(y) \right \}_{j =1}^B.$$
\item Construct preliminary joint DF-bands  $([L_k', U_k'])_{k \in \mK}$ for   $(F_k)_{k \in \mK}$  of level $p$
as
$$
[L'_k(y), U'_k(y)] = [ \hat{F}_k(y) \pm c(p) \hat s_k(y)], \quad y \in T, \quad k \in \mK.
$$
For each $k \in \mK$ impose the shape restrictions on $\hat F_k$, $L_k'$ and $U_k'$ as described in Appendix \ref{app:CBDF}.
\item Report $(I_k)_{k \in \mK}= ([L_k,U_k])_{k \in \mK}$ as p-level simultaneous DF-bands  for  $(F_k)_{k \in \mK}$.  Report $(I^\lto_k)_{k \in \mK}= ([U^\lto_k,L^\lto_k])_{k \in \mK}$ or the support-restricted version $(\tilde I^\lto_k)_{k \in \mK}
= (I_k^\lto \cap T_k)_{k \in \mK}$ as $p$-level simultanenous QF-bands  for  $(F^\lto_k)_{k \in \mK}$.  
\item Report $I^\lto_{\Delta(j,k)} = I^\lto_j  -  I^\lto_k$ or the support-restricted version
$\tilde I^\lto_{\Delta(j,k)}  = \tilde I^\lto_j  -  \tilde I^\lto_k$ as $p$-level simultaneous QE-bands for   $ F_j^{\leftarrow}  -  F_k^{\leftarrow}  $ for all $(j,k) \in \mK^2$.
\end{enumerate}
\end{algorithm}
\end{framed}

In  step (1) we bootstrap jointly all the estimators of the DFs. In our applications it
is important to obtain jointly the bootstrap draws of  these estimators because they are not independent. There are
multiple ways to obtain the bootstrap draws of $\hat{F}$. A generic
resampling procedure is the exchangeable bootstrap \citep{pw:93,vdV-W},
which recomputes $\hat{F}$ using sampling weights drawn independently from
the data. This procedure incorporates many popular bootstrap schemes as
special cases by a suitable choice of the distribution of the weights. For
example, the empirical bootstrap corresponds to multinomial weights, and the
weighted or Bayesian bootstrap corresponds to standard exponential weights.
Exchangeable bootstrap can also accommodate dependence or clustering in the
data by drawing the same weight for all the observations that belong to the
same cluster \citep{sc:97,cheng2013cluster}. For example, in the application
of Section \ref{subsec:doc} we draw the same weights for all the individuals
of the same household.

In the second step we estimate  pointwise standard errors. We use the bootstrap rescaled interquartile range because it is more robust than the bootstrap standard deviation in that it requires weaker conditions for consistency  \citep{chernozhukov+13inference}.   In the
third step, we compute, for each bootstrap draw, the weighted recentered
Kolmogorov-Smirnov maximal $t$-statistic over all distributions $F_{k}\left(
y\right) $ with $k\in \mathcal{K}$ and $y\in T$. The maximum over $k\in \mathcal{K}$ ensures
joint coverage of all the DFs.  Then we take the $p$%
-quantile of the bootstrap Kolmogorov-Smirnov statistics. This allows us, in
the fourth step, to construct preliminary DF-bands that jointly
cover all the DFs $\left( F_{k}\right) _{k\in \mathcal{K}}$ with probability 
$p$. We improve these bands by imposing the shape restrictions.

In the fifth step we invert the DF-bands to obtain QF-bands, as justified by Theorem \ref{theorem:bquant}. In the last step we
obtain the QE-bands by taking Minkowski
differences of the QF-bands, as justified by Theorem \ref{theorem:bqte}. If needed, we can
impose the support conditions in the last two steps.

The following corollary of Theorem \ref{theorem:bqte} provides theoretical
justification for Algorithm \ref{algorithm:qte}. To state the result, let $%
\ell ^{\infty }(\mathcal{Y})$ denote the metric space of bounded functions
from $\mathcal{Y}$ to $\mathbb{R}$ equipped with the sup-norm and $|\mathcal{K}|$ denote the
cardinality of the set $\mathcal{K}$.

\begin{corollary}[Validity of Algorithm 1]
\label{cor3} Suppose that the rescaled DF estimators $\{a_{n}(\hat{F}%
_{k}-F_{k})\}_{k\in \mathcal{K}}$ converge in law in $\ell ^{\infty }(%
\mathcal{Y})^{|\mathcal{K}|}$ to a Gaussian process $(G_{k})_{k\in \mathcal{K%
}}$, having zero mean and a non-degenerate variance function, for some
sequence of constants $a_{n}\rightarrow \infty $ as $n\rightarrow \infty $,
where $n$ is some index (typically the sample size). Suppose that a
bootstrap method can consistently approximate the limit law of $\{a_{n}(\hat{%
F}_{k}-F_{k})\}_{k\in \mathcal{K}}$, namely the distance between the law of $%
\{a_{n}(\hat{F}_{k}^{\ast }-\hat{F}_{k})\}_{k\in \mathcal{K}}$ conditional
on data, and that of $(G_{k})_{k\in \mathcal{K}}$, converges to zero in
probability as $n\rightarrow \infty $. The distance is the bounded Lipschitz
metric that metrizes weak convergence. Then, the confidence bands
constructed by Algorithm \ref{algorithm:qte} have the following covering
property: 
\begin{equation*}
\lim_{n\rightarrow \infty }\Pr (F_{k}\in I_{k},F_{k}^{\leftarrow }\in \tilde{%
I}_{k}^{\leftarrow },\Delta _{j,m}\in \tilde{I}_{\Delta (j,m)}^{\leftarrow };%
\text{ for all }(k,j,m)\in \mathcal{K}^{3})=p.
\end{equation*}
\end{corollary}

\textbf{Proof.} Lemma SA.1 of \citet{chernozhukov+13inference} implies that $%
\lim_{n\rightarrow \infty }\Pr (\cap _{k\in \mathcal{K}}\{F_{k}\in \lbrack
L_{k}^{\prime },U_{k}^{\prime }]\})=p$. The result then follows from Lemma %
\ref{lemma:shape}, Theorems \ref{theorem:bquant} and \ref{theorem:bqte}, and
Corollaries \ref{corollary:support} and \ref{corollary:support_qte}.\qed

Algorithm \ref{algorithm:qte} provides confidence bands that jointly cover
the DFs, the QFs, and the QE functions. If one is only interested in one
single QF, say $F_1^{\leftarrow}$, the corresponding QF-band
obtained from Algorithm \ref{algorithm:qte} can be conservative. This is
because we compute the maximal $t$-statistic over all distributions $%
\left(F_{k}\right)_{k\in \mathcal{K}} $ to ensure joint coverage, which is
not required if one is only interested in $F_1^{\leftarrow}$. Appendix \ref%
{app:algorithms_single_qf} provides a bootstrap algorithm that yields
an asymptotically similar QF-band for a single QF.

\begin{remark}[Validity of High-Level Conditions in Corollary \ref{cor3} for DR-based Estimators]
The high-level conditions in Corollary \ref{cor3} are satisfied by many estimators. In particular, Theorem 5.2 in \citet{chernozhukov+13inference} verifies these assumptions for the DR-based estimators that we use in the empirical applications in Section \ref{sec:applications}.\end{remark}

\section{Applications to Counterfactual Analysis Using Distribution Regression}

\label{sec:applications} In this section we apply our approach to two data
sets, corresponding to two common types of discrete outcomes.\footnote{The data and code in \texttt{R} \citep{R18} for the empirical analysis is available at \href{https://github.com/bmelly/discreteQ}{\texttt{https://github.com/bmelly/discreteQ}}.} In both cases
we use the distribution regression model and obtain QE as differences
between counterfactual distributions. For this reason, we first introduce
the specific methods and then present both empirical illustrations.

\subsection{Distribution Regression}

\label{subsec:dr}

In the absence of covariates, the empirical DF is a minimal sufficient
statistic for a non-parametric marginal DF. Distribution regression (DR)
generalizes this concept to a conditional DF like OLS generalizes the
univariate mean to the conditional mean function. The key, simple
observation underlying DR is that the conditional distribution of the
outcome $Y$ given the covariates $X$ at a point $y$ can be expressed as $F_{Y\mid X}(y\mid
x)={\mathrm{E}}[1\{Y\leq y\}\mid X=x]$. Accordingly, we can construct
a collection of binary response variables, which record the events that the
outcome $Y$ falls bellow a set of thresholds $T$, i.e., 
\begin{equation*}
1\{Y\leq y\},\quad y\in T,
\end{equation*}%
and use a binary regression model for each variable in this collection. This
yields the DR model: 
\begin{equation}
F_{Y\mid X}\left( y\mid x\right) =P(Y\leq y\mid X=x)=\Lambda _{y}\left(
B(x)^{\prime }\beta \left( y\right) \right) \text{,}
\end{equation}%
where $\Lambda _{y}(\cdot )$ is a known link function which is allowed to
change with the threshold level $y$; $B(x)$ is a vector of transformations
of $x$ with good approximating properties such as polynomials, B-splines,
and interactions; and $\beta \left( y\right) $ is an unknown vector of
parameters. Knowledge of the function $y\mapsto \beta (y)$ implies knowledge
of the distribution of $Y$ conditional on $X$. The DR model is flexible in
the sense that, for any given link function, we can approximate the
conditional DF arbitrarily well by using a rich enough set of
transformations of the original covariates $B(x)$. In the extreme case when $%
X$ is discrete and $B(x)$ is fully saturated, the estimated conditional
distribution is numerically equal to the empirical DF in each cell of $X$
for any monotonic link function. When $B(x)$ is not fully saturated, one can
choose a DF such as the normal or logistic as the link function to guarantee
that the model probabilities lie between 0 and 1.

DR nests a variety of classical models such as the Normal regression, the
Cox proportional hazard, ordered logit, ordered probit, Poisson regression,
as well as other generalized linear models. Example \ref{example:poisson}
shows the inclusion of the Poisson regression model which we use as a
benchmark in our first empirical application. In what follows we set $B(x)=x$
to lighten the notation without loss of generality.

\begin{example}
\label{example:poisson} Let $Y$ be a nonnegative integer-valued outcome and $%
X$ a vector of covariates. The Poisson regression model assumes that the
probability mass function of $Y$ conditional on $X$ is 
\begin{equation*}
f_{Y|X}\left( y\mid x\right) =\frac{\exp \left( x^{\prime }\beta \right)
^{y}\exp \left( -\exp \left( x^{\prime }\beta \right) \right) }{y!}\text{
for } y =\{0,1,2,...\}.
\end{equation*}%
The corresponding conditional distribution is: 
\begin{equation*}
F_{Y\mid X}\left( y\mid x\right) =\sum_{k=0}^{y}\frac{\exp \left( x^{\prime
}\gamma \right) ^{k}\exp \left( -\exp \left( x^{\prime }\beta \right)
\right) }{k!}=Q\left( y,\exp \left( x^{\prime }\beta \right) \right),
\end{equation*}%
where $Q$ is the incomplete gamma function. Thus, the Poisson regression can
be seen as a special case of a DR model with exponentiated incomplete gamma
link function, 
\begin{equation}  \label{eq:igf}
\Lambda _{y}\left( u \right) = Q\left( y,\exp u \right),
\end{equation}
and parameter function $y \mapsto \beta(y)$ that does not vary with $y$,
i.e. $\beta(y) = \beta$. 
The Poisson regression model therefore imposes strong homogeneity
restrictions on the effect of the covariates at different parts of the
distribution that are often rejected by the data (see, e.g., Section \ref%
{subsec:doc}). \qed
\end{example}

Assume that we have a sample $\{(Y_{i},X_{i}):i=1,...,n\}$ of $(Y,X)$. The
DR estimator of the conditional distribution is 
\begin{equation*}
\hat{F}_{Y|X}(y\mid x)=\Lambda _{y}(x^{\prime }\hat{\beta}(y)),\quad y\in T,
\end{equation*}%
where 
\begin{equation*}
\hat{\beta}(y) \in \arg \max_{b\in \mathbb{R}^{\text{dim}(X)}}\sum_{i=1}^{n}1%
\{Y_{i}\leq y\}\ln \left[ \Lambda _{y}\left( X_{i}{}^{\prime }b\right) %
\right] +1\{Y_{i}>y\}\ln \left[ 1-\Lambda _{y}\left( X_{i}{}^{\prime
}b\right) \right] .
\end{equation*}%
\citet{williams1972analysis} introduced DR in the context of ordered
outcomes. \citet{foresiperacchi95} applied this method to estimate the
conditional distribution of excess return evaluated at a finite number of
points. \citet{chernozhukov+13inference} extended %
\citet{williams1972analysis}'s definition to arbitrary outcomes and
established functional central limit theorems and bootstrap validity results
for DR as an estimator of the whole conditional distribution. One of the
main advantages of DR is that it not only accommodates continuous but also
discrete and mixed discrete continuous outcomes very naturally.

\subsection{Marginal and Counterfactual Distributions}

\label{subsec:counter}  We show how to utilize DR for causal inference in two empirical applications. In both applications there
are two groups: the treated and control units in the first application, and
the black and white children in the second application. We use DR to model
and estimate the conditional distribution of the outcome in each group at
each value of the covariates, that we denote by $F_{Y_{0}|X_{0}}(y\mid x)$
and $F_{Y_{1}|X_{1}}(y\mid x)$. The difference between these two
high-dimensional DFs is, however, difficult to convey. Instead, we integrate
these conditional distributions with respect to observed covariate
distributions and compare the resulting marginal distributions.

For instance, in the first application, the marginal distribution 
\begin{equation*}
F_{{\langle k\rangle }}(y):=\int F_{Y_{k}|X_{k}}(y\mid x)dF_{X}(x),
\end{equation*}%
where $F_{X}$ is the distribution of $X$ in the entire population including
the treated and control units, represents the distribution of a potential
outcome. When $k=1$, $F_{{\langle 1\rangle }}$ is the outcome distribution that would be observed if
every units were treated, and when $k=0$, $F_{{\langle 0\rangle }}$ is the outcome distribution if
every units were not treated. These two distributions are called
counterfactual, since they do not arise as distributions from any observable
population. They nevertheless  have a causal interpretation as distributions
of potential outcomes when the treatment is randomized conditionally on the
control variables $X$.  

Let $\hat{F}_{Y_{k}|X_{k}}$ denote the DR estimator of $F_{Y_{k}|X_{k}}$, $%
k\in \{0,1\}$. We estimate $F_{{\langle k\rangle }}$ by the plugging-in
rule, namely integrating $\hat{F}_{Y_{k}|X_{k}}$ with respect to the
empirical distribution of $X$ for treated and control units. For $k\in
\left\{ 0,1\right\} $,%
\begin{equation*}
\hat{F}_{{\langle k\rangle }}(y):=\frac{1}{n}\sum_{i=1}^{n}\hat{F}%
_{Y_{k}|X_{k}}(y\mid X_{i}).
\end{equation*}%
We then report the empirical QE function:%
\begin{equation*}
\hat{\Delta}(a):=\hat{F}_{{\langle 1\rangle }}^{\leftarrow }(a)-\hat{F}_{{%
\langle 0\rangle }}^{\leftarrow }(a),a\in \lbrack 0,1].
\end{equation*}

\citet{chernozhukov+13inference} derived joint functional central limit
theorems for $(\hat{F}_{{\langle 0\rangle }},\hat{F}_{{\langle 1\rangle }})$
and established bootstrap validity.
We can thus use the algorithms
in Section \ref{section:algorithms} to construct asymptotically valid
simultaneous confidence bands for the counterfactual QFs $(F_{{\langle
1\rangle }}^{\leftarrow },F_{{\langle 0\rangle }}^{\leftarrow })$ and the QE
function $\Delta =F_{{\langle 1\rangle }}^{\leftarrow }-F_{{\langle 0\rangle 
}}^{\leftarrow }$.

\begin{remark}[Continuous covariates]
The proposed approach can also be used to analyze the effect of continuous
covariates. For instance, we can compare the status quo QF with the QF that
we would observe if everyone received $\Delta d$ additional units of the
continuous covariate of interest $D$, e.g. $\Delta d = 1$ for a unitary
increase. Formally, assume that we are interested in the effect of a
continuous variable $D$ on the outcome $Y$ while controlling for a vector of
covariates $X$. We can define the counterfactual distribution 
\begin{equation*}
F_{\langle \Delta d\rangle}(y):=\int F_{Y|D,X}(y\mid d+\Delta
d,x)dF_{D,X}(d,x)
\end{equation*}%
and the QE function $F_{\langle \Delta d\rangle}^{\leftarrow }(a)-F_{\langle
0 \rangle}^{\leftarrow }(a)$, where $F_{\langle 0 \rangle}$ is the marginal
(status quo) distribution of $Y$. This experiment can be interpreted as an
unconditional quantile regression. Also in this case, our methods provide
valid confidence bands for the counterfactual quantile and QE functions.%
\qed
\end{remark}

\subsection{Insurance coverage and health care utilization}

\label{subsec:doc}

Our first application illustrates the construction of confidence bands
using data from the Oregon health insurance experiment. In 2008, the state
of Oregon initiated a limited expansion of its Medicaid program for
uninsured low-income adults by offering insurance coverage to the lottery
winners from a waiting list of 90,000 people (see \href{http://www.nber.org/oregon/}%
{www.nber.org/oregon} for details). This experiment constitutes a unique
opportunity to study the impact of insurance by means of a large-scale
randomized controlled trial %
\citep[e.g.,][]{finkelstein+12,baicker+13,baicker+14,taubman+14}.

We investigate the impact of insurance coverage on health care utilization
as analyzed in \citet[][Section V]{finkelstein+12} using a publicly
available dataset \citep{finkelstein+12data}. The data are available via: \href{http://www.nber.org/oregon/4.data.html}{%
http://www.nber.org/oregon/4.data.html}. Detailed information about the
dataset and descriptive statistics are available in \citet{finkelstein+12}
and the corresponding online appendix. We focus on one count outcome $Y$:
the number of outpatient visits in the last six months, which was elicited
via a large mail survey. After excluding individuals with missing
information in any of the variables used in the analysis, the resulting
sample consists of 23,441 observations. The top histogram in Figure \ref%
{figure:histograms} illustrates the discrete nature of our dependent
variable. Almost 40\% of the outcomes are zeros, more than 90\% of the mass
is concentrated between zero and five, but a few people have a greater
number of visits.

\citet{finkelstein+12} find a positive effect of winning the lottery on the
number of outpatient visits.\footnote{They label these effects intention-to-treat (ITT) effects and also report
local average treatment effects (LATE) estimated using IV regressions. In
this section, we focus on ITT effects.} Their results are based on ordinary
least squares (OLS) regressions, where the covariates $X$ include household
size, indicators for the survey wave, and interactions of the household size
indicators and the survey wave. Although individuals were chosen randomly,
these covariates are included as controls because the entire household for
any selected individual became eligible to apply for insurance and the
fraction of treated individuals varies across survey waves. We complement
their findings by looking at the whole distribution of the number outpatient
visits. We first estimate the conditional outcome distributions separately
for the lottery winners and losers via Poisson regression and DR. For DR, we
use the exponentiated incomplete gamma link in \eqref{eq:igf} such that DR
nests the Poisson regression as an exact special case. As explained in
Section \ref{subsec:counter}, we integrate the conditional outcome
distributions with respect to the covariate distribution for both lottery
winners and losers to obtain estimates of the counterfactual distributions $%
F_{\left\langle 1\right\rangle }$ and $F_{\left\langle 0\right\rangle }$.

\begin{figure}[tbph]
\begin{center}
\includegraphics[height=7.5in]{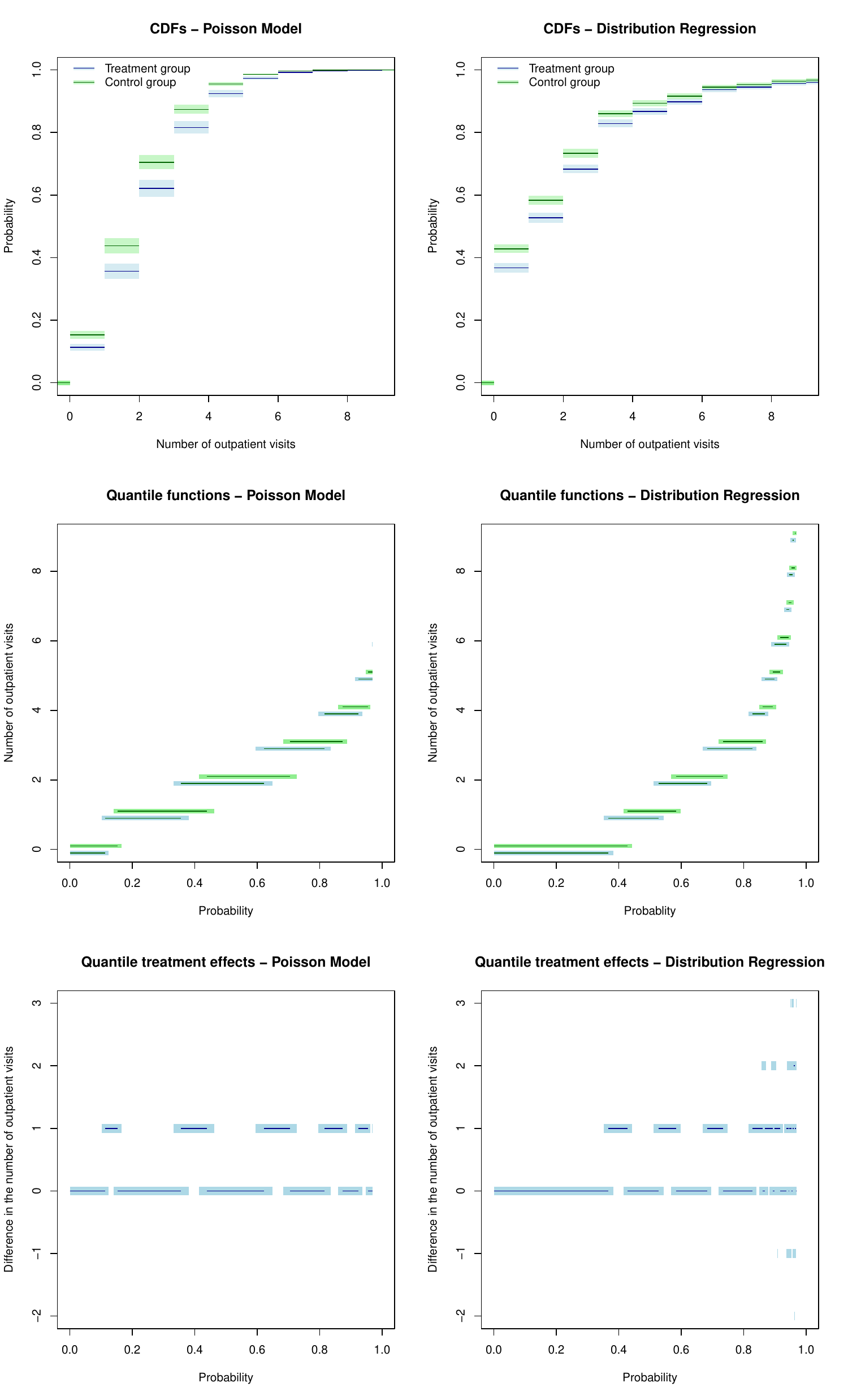}
\end{center}
\caption{Effect of insurance coverage on the number of outpatient visits in
the last six months. DFs, QFs, and QTE estimated by Poisson
regression and DR including support restricted 95\%
confidence bands. The lines of the QF for the control group
are slightly shifted upward to avoid overlapping with the QF
for the treatment group.}
\label{figure:results_oregon_doc}
\end{figure}

The top panel of Figure \ref{figure:results_oregon_doc} displays the DFs $%
\hat{F}_{\left\langle 1\right\rangle }$ and $\hat{F}_{\left\langle
0\right\rangle }$ estimated by the Poisson regression and DR. The
corresponding QFs $\hat{F}_{\left\langle 1\right\rangle }^{\leftarrow }$ and 
$F_{\left\langle 0\right\rangle }^{\leftarrow }$ are displayed in both
middle panels. Finally, the estimated QE functions, $\hat{F}_{{\langle
1\rangle }}^{\leftarrow }-\hat{F}_{{\langle 0\rangle }}^{\leftarrow }$, are
plotted in the bottom panels. In all cases, the figure also shows 95\%
simultaneous confidence bands, constructed using Algorithm  \ref{algorithm:qte} with $B=1,000$ Bayesian bootstrap
draws that take into account the possible clustering of the observations at
the household level. 
Reflecting the discrete nature of our outcome variables, we impose the
support restrictions $T_0 = T_1 =\{0,1,\ldots \}$.

A comparison between the Poisson and DR results reveals striking
differences. The Poisson model predicts a much lower mass at zero and a much
thinner upper tail of the distribution for both groups. Indeed, these
differences are statistically significant as the Poisson and DR simultaneous
DF-bands  and QF-bands do not overlap for a large part of the
support. A formal test rejects the equality of these distributions with a
p-value below $0.001$. Since the DR model with exponentiated incomplete
gamma link nests the Poisson model, we conclude that the Poisson model is
rejected by the data. For this reason, we focus the discussion on the DR
results.

The  QE-band do not fully cover the zero-line
and thus we can reject the null hypothesis that winning the lottery has no
effect on the number of outpatient visits. We can also reject the hypothesis
that $F_{\left\langle 0\right\rangle }$ first-order stochastically dominates 
$F_{\left\langle 1\right\rangle }$ because the band for $F_{\left\langle
0\right\rangle }^{\leftarrow}$ is strictly below the band for $%
F_{\left\langle 1\right\rangle }^{\leftarrow}$ at some probability indexes. However,
we cannot reject the opposite hypothesis. In other words, at no quantile
index the confidence band contains strictly negative effects while at some
probability indexes it contains strictly positive effects.

Health economists distinguish between the treatment effect on the extensive
(whether to see a doctor) and intensive (the number of visits given at least
one) margins. The first effect is easy to estimate: the probability of not
seeing a doctor decreased significantly from 43\% to 37\% with the
treatment. The effect on the intensive margin is more difficult to gauge
because we do not observe both potential outcomes for any individual. If we
assume that the individuals induced to see a doctor by the insurance
coverage are not seriously sick and visit the doctor only once, then the
effect on the intensive margin can also be seen in Figure \ref%
{figure:results_oregon_doc}: the effect from 0 to 1 visit represents the
effect on the extensive margin and the effect on the rest of the
distribution represents the effect on the intensive margin. Both effects are
statistically significant. We note in particular that the quantile
differences do not vanish at the top of the distribution.

The assumption made to justify this interpretation may be too strong and
lead to an overestimation of the effect on the intensive margin. For
instance, the doctor may find a serious problem and schedule other visits.
Following \citet{zhang2003estimation} and \citet{angrist2006long}, we can
bound the effect on the intensive margin from below by assuming that
patients who see a doctor anyway visit their doctor at least as often as
patients who see a doctor only if insured. Under this weaker assumption, the
effect on the intensive margin is bounded from below by the QE function
obtained by keeping only observations with at least one visit. We also find
a positive treatment effect with this method, which reinforces the evidence
of a positive effect among the existing users.

\subsection{Racial differences in mental ability of young children}

\label{subsection:testscores} As a second application, we reanalyze the
racial IQ test score gap examined in \citet{fryer2013testing}. We use data from
the US Collaborative Perinatal Project (CPP). These data contain information
on children from 30,002 women who gave birth in 12 medical centers between
1959 and 1965. Our main outcomes of interest are the standardized test
scores at the ages of eight months (Bayley Scale of Infant Development) and
seven years (both Stanford-Binet and Wechsler Intelligence Test). In
addition to the test score measures, the dataset contains a rich set of
background characteristics for the children, $X$, including information on
age, gender, region, socioeconomic status, home environment, prenatal
conditions, and interviewer fixed effects. \citet{fryer2013testing} provide a
comprehensive description of the dataset and extensive descriptive
statistics.

A key feature of the test scores is the discrete nature of their
distribution. We observe only 76 and 128 different values for the
standardized test scores at the ages of eight months and seven years,
respectively. The middle and bottom panels of Figure \ref{figure:histograms}
present the corresponding histograms. Note that each bar corresponds to
exactly one value. For instance, at eight months, almost 12\% of the
observations have exactly the same score and 60\% of the observations have
one of the most frequent six values. This is a common feature of test
scores, which are necessarily discrete because they are based on a finite
number of questions.

To gain a better understanding of the causes of the observed black-white
test score gap, we provide a distributional decomposition into explained and
unexplained parts by observable background characteristics. Let $F_{{\langle
W|W\rangle }}$ and $F_{{\langle B|B\rangle }}$ represent the observed test
score DFs for white and black children, and $F_{\langle W|B\rangle }$
represents the counterfactual DF of test scores that would have prevailed for
white children had they had the distribution of background characteristics
of black children, $F_{X_{B}}$, namely, 
\begin{equation}
F_{{\langle W|B\rangle }}(y):=\int F_{Y_{W}|X_{W}}(y\mid x)dF_{X_{B}}(x).
\label{counter_testscores}
\end{equation}%
With this counterfactual test score distribution it is possible to decompose
the quantiles of the observed black-white test score gap into 
\begin{equation}
F_{\langle W\mid W\rangle }^{\leftarrow }-F_{\langle B\mid B\rangle
}^{\leftarrow }=[F_{\langle W\mid W\rangle }^{\leftarrow }-F_{\langle W\mid
B\rangle }^{\leftarrow }]+[F_{\langle W\mid B\rangle }^{\leftarrow
}-F_{\langle B\mid B\rangle }^{\leftarrow }].
\end{equation}%
where the first term in brackets corresponds to the composition effect due
to differences in observable background characteristics and the second term
is the unexplained difference.

We estimate $F_{\langle W\mid W\rangle }$ and $F_{\langle B\mid B\rangle }$
by the empirical test score distributions for white and black children,
respectively. We estimate the counterfactual distribution $F_{\langle W\mid
B\rangle }$ by the sample analog of \eqref{counter_testscores} replacing $%
F_{Y_{W}|X_{W}}$ by the DR estimator for white children, and $F_{X_B}$ by
the empirical distribution of $X$ for black children. We use the logistic
link function for the DR, but the results using the linear link function or
the normal link function are similar.

\begin{figure}[tbph]
\begin{center}
\includegraphics[height=5in]{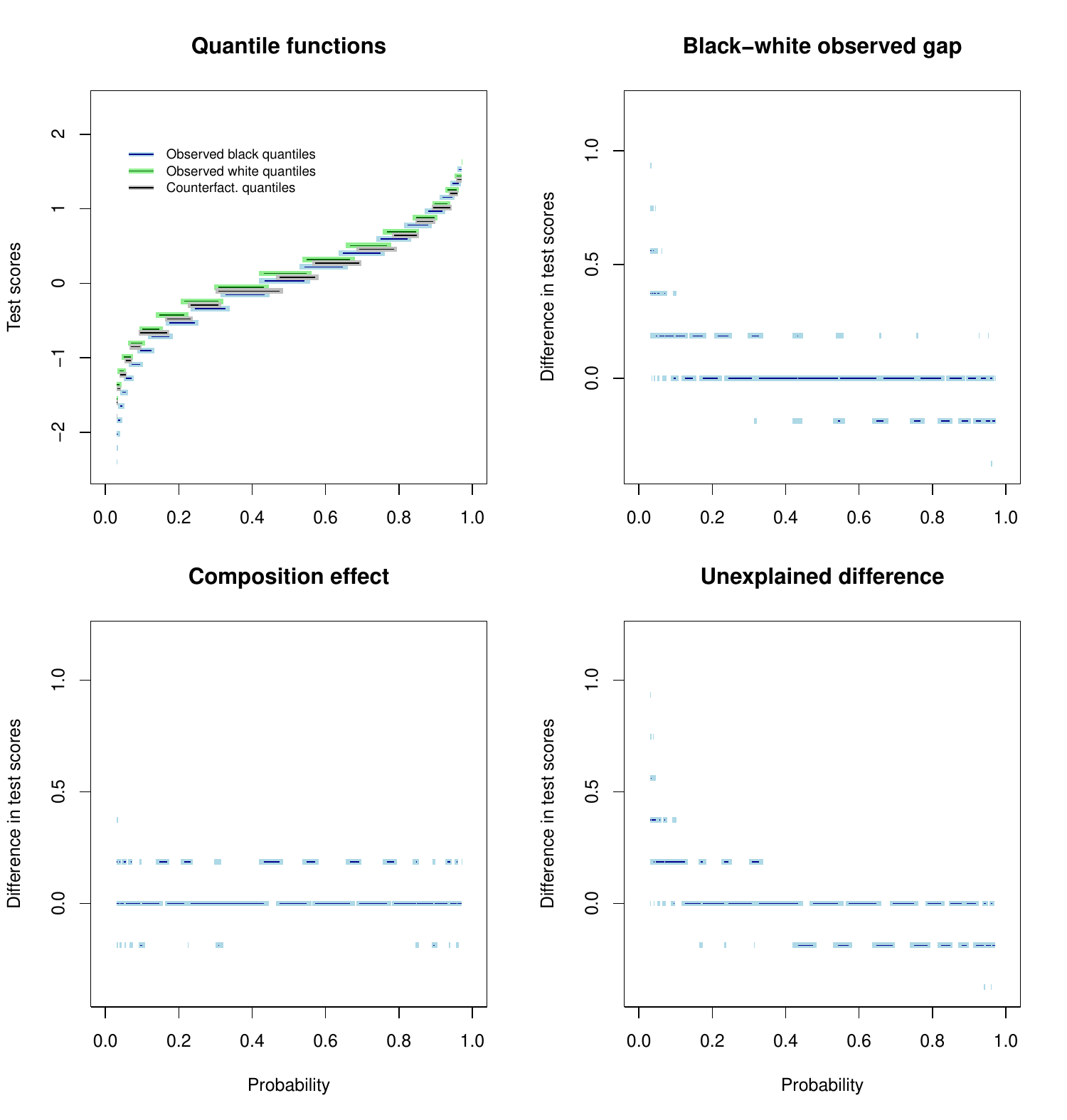}
\end{center}
\caption{Decomposition of observed racial differences in mental ability of
young children; results for eight months old children. QFs,
raw difference, composition effect, and unexplained difference including
support restricted 95\% confidence bands. The QF lines have
been slightly shifted vertically to avoid overlap.}
\label{figure:results_testscores_9m}
\end{figure}

\begin{figure}[tbph]
\begin{center}
\includegraphics[height=5in]{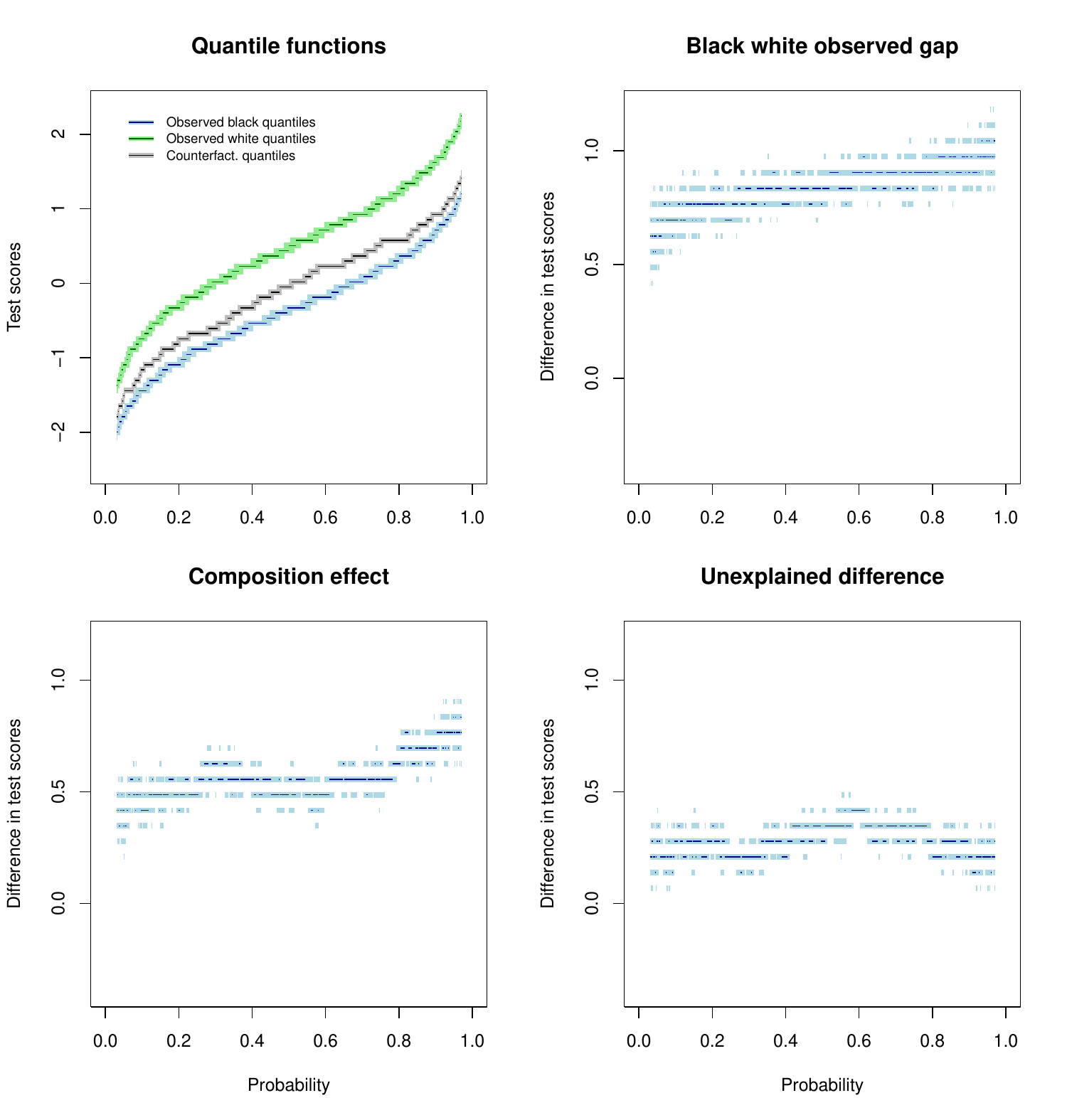}
\end{center}
\caption{Decomposition of observed racial differences in mental ability of
young children; results for seven year old children. QFs, raw
difference, composition effect, and unexplained difference including support
restricted 95\% confidence bands. }
\label{figure:results_testscores_7y}
\end{figure}


Figures \ref{figure:results_testscores_9m} and \ref%
{figure:results_testscores_7y} report the results for the eight months and
seven years outcomes, respectively. The first panels show the observed and counterfactual
QFs, $F_{\langle W\mid W\rangle }^{\leftarrow }$, $F_{\langle B\mid B\rangle
}^{\leftarrow }$ and $F_{\langle W\mid B\rangle }^{\leftarrow }$. The second
panels show the difference between the observed QFs, $F_{\langle W\mid
W\rangle }^{\leftarrow }-F_{\langle B\mid B\rangle }^{\leftarrow }$. The
third and fourth panels decompose these observed differences into the
composition effect ($F_{\langle W\mid W\rangle }^{\leftarrow }-F_{\langle
W\mid B\rangle }^{\leftarrow }$) and the unexplained component ($F_{\langle
W\mid B\rangle }^{\leftarrow }-F_{\langle B\mid B\rangle }^{\leftarrow }$).
The point estimates are shown with their respective 95\% simultaneous
confidence bands constructed using Algorithm  \ref%
{algorithm:qte} with $B=1,000$ Bayesian bootstrap draws. The bands impose
the restrictions that the supports of the test scores correspond to the
observed values in the sample.

For eight-month-old children, we find very small differences between the
test score distributions of black and white children. The black-white gap is
positive at the lower tail and is mainly due to unobserved characteristics.
While these effects are statistically significant, they are so small in
magnitude that they should not worry any policy maker. The composition
effect is very small, probably simply because there was no difference to
explain to begin with.

The results are completely different for seven-year-old children. We find a
large and statistically significant positive raw black-white gap. A formal
test based on the uniform bands rejects the null hypothesis of a zero or a
negative racial test score gap at all quantiles. 
The estimated QE function is increasing in the probability index ranging from
below $0.6$ standard deviation units at the lower tail up to over one
standard deviation unit at the upper tail of the distribution. The quantile
differences at the tails substantially differ from the mean difference of $%
0.85$ standard deviation units reported in \citet{fryer2013testing}. In fact,
we can formally reject the null hypothesis of a constant raw test score gap
across the distribution because we can not draw a horizontal line at any
value of the difference of test scores, which is covered by the confidence
band of the QE function at all probability indexes.

Our decomposition analysis shows that about two third of this gap can be
explained by differences in the distribution of observable characteristics.
Nevertheless, the remaining unexplained difference is significant, both in
economic and in statistical terms. Looking at the QE function, we can see
that there is substantial effect heterogeneneity along the distribution.
Interestingly, the increase in the test score gap at the upper quantiles can
be fully explained by differences in background characteristics between
black and white children. The resulting unexplained difference is maximized
in the center of the distribution. Finally, our simultaneous confidence
bands allow for testing several interesting hypothesis' about the whole QE
function. For instance, we can reject the null hypothesis that the
composition effect and the unexplained difference are zero, negative, or
constant at all quantiles but we cannot reject that they are positive
everywhere.


\bibliographystyle{elsarticle-harv}
\bibliography{bibliography}

\newpage

\begin{appendix}
\setcounter{page}{1}

\section{Imposing Monotonicity and Range Restrictions on Estimates and Confidence Bands for Distribution Functions}
\label{app:CBDF}

In many applications the point estimates $\hat{F}$ and interval estimates $%
[L^{\prime },U^{\prime }]$ for the target distribution $F$ do not satisfy
logical monotonicity or range restrictions, namely they do not take values
in the set $\mathbb{D}$ defined in Section \ref{sec:bands}. Given such an ordered triple $L^{\prime }\leq \hat{F%
}\leq U^{\prime }$, we can always transform it into another ordered triple $%
L\leq \check{F}\leq U$ that obeys the logical monotonicity and shape
restrictions. For example, we can set 
\begin{equation}
\check{F}=\mathcal{S}(\hat{F}),\quad L=\mathcal{S}(L^{\prime }),\quad U=%
\mathcal{S}(U^{\prime }),  \label{construct}
\end{equation}%
where $\mathcal{S}$ is the shaping operator that given a function $y \mapsto
f(y)$ yields a mapping $y\mapsto \mathcal{S}(f)(y)\in \mathbb{D}$ with 
\begin{equation*}
\mathcal{S}(f)=\mathcal{M}(0\vee f\wedge 1),
\end{equation*}%
where the maximum and minimum are taken pointwise, and $\mathcal{M}$ is the
rearrangement operator that given a function $f:\mathcal{Y} \mapsto \lbrack
0,1]$ yields a map $y \mapsto \mathcal{M}(f)(y)\in \mathbb{D}.$ 
Other monotonization operators, such as the projection on the set of weakly
increasing functions, can also be used, as we remark further below.

The \textit{rearrangement operator} is defined as follows. Let $T$ be a
countable subset of $\mathcal{Y}$. In our leading case where $f$ is the
distribution function of a discrete random variable $Y$, we can choose $T$
as the support of $Y$ and extend $f$ to $\mathcal{Y}$ by constant
interpolation, yielding a step function as the distribution of $Y$ on $%
\mathcal{Y}$. If $f$ is a distribution function of a continuous or mixed
random variable $Y$, we can set $T$ as a grid of values covering the support
of $Y$ where we evaluate $f$ and extend $f$ to $\mathcal{Y}$ by linear
interpolation. Given a $f:T\mapsto \lbrack 0,1]$, we first consider $%
\mathcal{M}f$ as a vector of sorted values of the set $\{f(t):t\in T\}$,
where the sorting is done in a non-decreasing order. Since $T$ is an ordered
set of the same cardinality as $\mathcal{M}f$, we can assign the elements of 
$\mathcal{M}f$ to $T$ in one-to-one manner: to the $k$-th smallest element
of $T$ we assign the $k$-th smallest element of $\mathcal{M}f$. The
resulting mapping $t\mapsto \mathcal{M}f(t)$ is the rearrangement operator.
We can extend the rearranged function $\mathcal{M}f$ to $\mathcal{Y}$ by
constant or linear interpolation as we describe above.

The following lemma shows that shape restrictions \textit{improve} the
finite-sample properties of the estimators and confidence bands.

\begin{lemma}[Shaping Improves Point and Interval Estimates]
\label{lemma:shape} The shaping operator $\mathcal{S}$

\begin{itemize}
\item[(a)] is weakly contractive under the max distance: 
\begin{equation*}
\| \mathcal{S}(A) - \mathcal{S}(B)\|_{\infty} \leq \| A - B\|_\infty, \quad 
\text{ for any $A$, $B$: $T \to [0,1]$, }
\end{equation*}

\item[(b)] is shape-neutral, 
\begin{equation*}
\mathcal{S}(F) = F \text{ for any } F \in \mathbb{D},
\end{equation*}

\item[(c)] and preserves the partial order: 
\begin{equation*}
A \leq B \ \implies \mathcal{S}(A) \leq \mathcal{S}(B), \quad \text{ for any 
$A$, $B$: $T \to [0,1]$. }
\end{equation*}
\end{itemize}

Consequently,

\begin{enumerate}
\item the re-shaped point estimate constructed via (\ref{construct}) is
weakly closer to $F$ than the initial estimate under the max distance: 
\begin{equation*}
\| \check F - F\|_\infty \leq \|\hat F- F\|_\infty,
\end{equation*}

\item the re-shaped confidence band constructed via (\ref{construct}) has
weakly greater coverage than the initial confidence band: 
\begin{equation*}
\Pr( L^{\prime }\leq F \leq U^{\prime }) \leq \Pr( L \leq F \leq U),
\end{equation*}

\item and the re-shaped confidence band is weakly shorter than the original
confidence bands under the max distance, 
\begin{equation*}
\| U - L \|_\infty \leq \|U^{\prime }- L^{\prime }\|_\infty.
\end{equation*}
\end{enumerate}
\end{lemma}

\textbf{Proof}. The result follows from \citet{CFG09}. \qed

The band $[L,U]$ is therefore weakly better than the original band $%
[L^{\prime },U^{\prime }]$, in the sense that coverage is preserved while
the width of the confidence band is weakly shorter.

\begin{remark}[Isotonization is Another Option]
An alternative to the rearrangement is the isotonization, which projects a
given function on the set of weakly increasing functions that map $T$ 
to $[0,1]$. This also has the improving properties stated in Lemma \ref%
{lemma:shape}. In fact any convex combination between isotonization and
rearrangement has the improving properties stated in Lemma \ref{lemma:shape}%
. \qed
\end{remark}

\begin{remark}[Shape Restrictions on Confidence Bands by Intersection]
An alternative way of imposing shape restrictions on the confidence band, is
to intersect the initial band $[L^{\prime },U^{\prime }]$ with $\mathbb{D}$.
That is,
we simply set 
\begin{equation*}
[L^I,U^I] = \mathbb{D} \cap [L^{\prime }, U^{\prime }] = \{ w \in \mathbb{D} : L^{\prime }(y) \leq w(y) \leq U^{\prime }(y), \quad \forall y \in \mathcal{Y} \}.
\end{equation*}
Thus, $U^I$ is the greatest nondecreasing minorant of $0 \vee U^{\prime
}\wedge 1$ and $L^I$ is the smallest nodecreasing majorant of $0 \vee
L^{\prime }\wedge 1$. This approach gives the tightest confidence bands, in
particular 
\begin{equation*}
[L^I,U^I] \subseteq [L,U].
\end{equation*}
However, this construction might be less robust to misspecification than the
rearrangement. For example, imagine that the target function $F$ is not
monotone, i.e. $F \not \in \mathbb{D}$. This situation might arise when $F$
is the probability limit of some estimator $\hat F$ that is inconsistent for
the DF due to misspecification. If the confidence band $[L^{\prime
},U^{\prime }]$ is sufficiently tight, then we can end up with an empty
intersection band, $[L^I,U^I] = \emptyset$. By contrast $[L,U]$ is non-empty
and covers the reshaped target function $F^*= \mathcal{S}(F) \in \mathbb{D}$.%
\qed
\end{remark}

\section{Bootstrap Algorithms for Confidence Bands for Single Quantile Functions}
\label{app:algorithms_single_qf}

If one is only interested in a single QF $F^\leftarrow$, the QF-band constructed based on Algorithm \ref{algorithm:qte} will generally be conservative. Here, we provide an algorithm that provides asymptotically similar (non-conservative) uniform confidence bands that jointly cover the DF, $F$, and the corresponding QF, $F^{\leftarrow }$.

\begin{framed}
\begin{algorithm}[Bootstrap Algorithm for Single QF-Band]
\text{ }
\label{algorithm:cdf_qf}
\begin{enumerate} 
\item Obtain  many bootstrap draws  of the estimator   $\hat{F}$, 
$$\hat{F}^{*(j)}, \quad  j=1, \ldots, B$$
where the index $j$ enumerates the bootstrap draws and $B$ is the number of bootstrap draws (e.g., $B = 1,000$). 

\item For each $y$ in $T$,  compute the robust standard error of $\hat{F}(y)$, 
$$
\hat s(y) =  (\hat Q(.75,y)- \hat Q(.25,y))/(\Phi^{-1}(.75) - \Phi^{-1}(.25)),$$
where  $\hat Q(\alpha,y)$ denotes the empirical $\alpha$-quantile of the bootstrap sample $(\hat{F}^{*(j)}(y))_{j=1}^B$, 
and $\Phi^{-1}$ denotes the inverse of the standard normal distribution.

\item Compute the critical value 
$$c(p) = \text{ $p$-quantile of  } \left \{ \max_{y \in T} |\hat{F}(y)^{*(j)} - \hat{F}(y)|/\hat s(y)\right \}_{j =1}^B.$$
\item Construct a preliminary DF-band $[L', U']$ for   $F$  of level $p$
via:
$
[L'(y), U'(y)] = [ \hat{F}(y) \pm c(p) \hat s(y)]$ for each  $y \in T.$ Impose the shape restrictions on $\hat F$, $L'$ and $U'$ as described in Appendix \ref{app:CBDF}.
Report $I= [L,U]$ as a $p$-level DF-band  for  $F$. 

\item  Report the inverted band $I^\lto= [U^\lto,L^\lto]$  or  support restricted inverted band $\tilde I^\lto =I^\lto \cap T$   as a $p$-level QF-band  for  $F^\lto$

\end{enumerate}
\end{algorithm}
\end{framed}

The following corollary of Theorem \ref{theorem:bquant} provides theoretical justification for Algorithm \ref{algorithm:cdf_qf}.
\begin{corollary}[Validity of Algorithm \ref{algorithm:cdf_qf}]
\label{corollary:validity_algorithm1} Suppose that the rescaled DF estimator 
$a_{n}(\hat{F}-F)$ converges in law in $\ell ^{\infty }(\mathcal{Y})$ to a
Gaussian process $G$, having zero mean and a non-degenerate variance
function, for some sequence of constants $a_{n}\rightarrow \infty $ as $%
n\rightarrow \infty $, where $n$ is some index (typically the sample size).
Suppose that a bootstrap method can consistently approximate the limit law
of $a_{n}(\hat{F}-F)$, namely the distance between the law of $a_{n}(\hat{F}%
^{\ast }-\hat{F})$ conditional on data, and that of $G$, converges to zero
in probability as $n\rightarrow \infty $. The distance is the bounded
Lipschitz metric that metrizes weak convergence. Then, 
\begin{equation*}
\lim_{n\rightarrow \infty }\Pr (F\in I,F^{\leftarrow }\in \tilde{I}%
^{\leftarrow })=p.
\end{equation*}
\end{corollary}

\textbf{Proof.} Lemma SA.1 of \citet{chernozhukov+13inference} implies that $%
\lim_{ n \to \infty} \Pr( F \in [L^{\prime },U^{\prime }]) =p$. The result
then follows from Lemma \ref{lemma:shape}, Theorem \ref{theorem:bquant} and
Corollary \ref{corollary:support}. \qed

\section{Simulation Study}\label{app:sim}
This section presents simulation evidence on the finite sample performance of our bands. To keep the simulations computationally tractable we analyze a setup without covariates. We generate two independent random samples $\{Y_{1i}\}_{i=1}^{n}$ and $\{Y_{0i}\}_{i=1}^{n}$ for the treated and control outcomes, respectively. The estimators of the DFs $\hat{F}_{Y_{0}}$ and  $\hat{F}_{Y_{1}}$  are simply the empirical distribution functions in the respective samples. We perform $5000$ simulations and let the sample size $n\in {\{400,1,600,6,400\}}$  vary in order to examine the convergence of the coverage rates with respect to the sample size.
 We consider the problem of constructing uniform confidence bands that cover (i) a single QF: either $F^{\leftarrow}_{Y_{1}}$ or $F^{\leftarrow}_{Y_{0}}$, (ii) simultaneously both DFs, both QFs and the QE function:  $F_{Y_{0}}$,  $F_{Y_{1}}$, $F^{\leftarrow}_{Y_{0}}$, $F^{\leftarrow}_{Y_{1}}$ and $F^{\leftarrow}_{Y_{1}}-F^{\leftarrow}_{Y_{0}}$, (iii) only the QE function: $F^{\leftarrow}_{Y_{1}}-F^{\leftarrow}_{Y_{0}}$. The confidence bands for a coverage
of type (i) are constructed based on Algorithm \ref{algorithm:cdf_qf} while the bands for a coverage of type (ii) or (iii) are constructed based on Algorithm \ref{algorithm:qte}.
We consider three confidence levels $p\in {\{0.9,0.95,0.99\}}$.

We consider two families of distributions: a count variable similar to the outcome in the first application and an ordered variable similar to the outcome in the second application.
 In the first case,  $Y_{1i}$ is distributed Poisson  with parameter   $\lambda=3$ and  $Y_{0i}$ is  distributed Poisson with $\lambda \in \{3,2.75,2.5\}$. Since the support of the Poisson distribution is unbounded, we estimate the QFs and QE functions for $a\in [0.1,0.9]$ and invert the bands for the DF over the part of the support that is relevant for the range of quantiles considered. Table \ref{table:performance.count} displays the empirical coverage rate of the true functions. We report the coverage rate of the DFs and QFs in a single column because they are numerically equal by construction.  We also  provide the empirical probability to reject the null hypothesis that  $F^{\leftarrow}_{Y_{1}}=F^{\leftarrow}_{Y_{0}}$. This allows us to measure the empirical size in the first panel (where this hypothesis is satisfied) and the empirical power in the other panels.

\newcolumntype{C}[1]{>{\centering\arraybackslash}m{#1}}
\begin{table}[ht]
\caption{Performance of the uniform bands for count outcomes}
\label{table:performance.count}
\centering
\begin{tabular}{r C{0.5cm} C{1.6cm} C{1.6cm} C{1.6cm} C{1.6cm} C{3cm}}
  \toprule
  \midrule
\multicolumn{1}{c}{$n$} & $p$ & \multicolumn{4}{c}{Empirical coverage probability for} & Prob. to reject\\
 & & $F_0$, $F^{\leftarrow}_0$ & $F_1$, $F^{\leftarrow}_1$ & all fct. & $F^{\leftarrow}_1-F^{\leftarrow}_0$  & $F^{\leftarrow}_1=F^{\leftarrow}_0$ 
\\ \hline
\multicolumn{7}{c}{Design 1: $Y_{0}\sim Poisson(3)$ and $Y_{1}\sim Poisson(3)$} \\
  400 & 0.99 & 0.99 & 0.99 & 0.99 & 1.00 & 0.00 \\ 
  400 & 0.95 & 0.96 & 0.96 & 0.97 & 1.00 & 0.00 \\ 
  400 & 0.90 & 0.92 & 0.92 & 0.92 & 1.00 & 0.00 \\ 
  1,600 & 0.99 & 0.99 & 0.99 & 0.99 & 1.00 & 0.00 \\ 
  1,600 & 0.95 & 0.96 & 0.96 & 0.96 & 1.00 & 0.00 \\ 
  1,600 & 0.90 & 0.92 & 0.91 & 0.92 & 1.00 & 0.00 \\ 
  6,400 & 0.99 & 0.99 & 0.99 & 0.99 & 1.00 & 0.00 \\ 
  6,400 & 0.95 & 0.96 & 0.95 & 0.95 & 1.00 & 0.00 \\ 
  6,400 & 0.90 & 0.91 & 0.91 & 0.90 & 1.00 & 0.00 \\ 
\hline
\multicolumn{7}{c}{Design 2: $Y_{0}\sim Poisson(3)$ and $Y_{1}\sim Poisson(2.75)$} \\
  400 & 0.99 & 0.99 & 0.99 & 1.00 & 1.00 & 0.00 \\ 
  400 & 0.95 & 0.96 & 0.96 & 0.96 & 0.98 & 0.03 \\ 
  400 & 0.90 & 0.92 & 0.92 & 0.92 & 0.94 & 0.07 \\ 
  1,600 & 0.99 & 0.99 & 0.99 & 0.99 & 0.99 & 0.20 \\ 
  1,600 & 0.95 & 0.96 & 0.95 & 0.95 & 0.96 & 0.48 \\ 
  1,600 & 0.90 & 0.92 & 0.91 & 0.91 & 0.91 & 0.65 \\ 
  6,400 & 0.99 & 0.99 & 0.99 & 0.99 & 0.99 & 1.00 \\ 
  6,400 & 0.95 & 0.96 & 0.96 & 0.95 & 0.95 & 1.00 \\ 
  6,400 & 0.90 & 0.91 & 0.91 & 0.91 & 0.91 & 1.00 \\   \hline
\multicolumn{7}{c}{Design 3: $Y_{0}\sim Poisson(3)$ and $Y_{1}\sim Poisson(2.5)$} \\
  400 & 0.99 & 0.99 & 0.99 & 0.99 & 0.99 & 0.16 \\ 
  400 & 0.95 & 0.96 & 0.96 & 0.96 & 0.97 & 0.47 \\ 
  400 & 0.90 & 0.92 & 0.92 & 0.92 & 0.93 & 0.65 \\ 
  1,600 & 0.99 & 0.99 & 0.99 & 0.99 & 0.99 & 1.00 \\ 
  1,600 & 0.95 & 0.96 & 0.96 & 0.96 & 0.96 & 1.00 \\ 
  1,600 & 0.90 & 0.92 & 0.91 & 0.92 & 0.92 & 1.00 \\ 
  6,400 & 0.99 & 0.99 & 0.99 & 0.99 & 0.99 & 1.00 \\ 
  6,400 & 0.95 & 0.96 & 0.95 & 0.95 & 0.95 & 1.00 \\ 
  6,400 & 0.90 & 0.91 & 0.92 & 0.91 & 0.91 & 1.00 \\ 
   \midrule
   \bottomrule
   \multicolumn{7}{p{11cm}}{\scriptsize{\it Notes:} Based on $5,000$ simulations.}
\end{tabular} 
\end{table}

The empirical coverage rates of the bands for a single QF (in the third and fourth columns of Table \ref{table:performance.count}) as well as the coverage rate for both DFs, both QFs and the QE function (in the fifth column) confirms the theoretical results in corollaries \ref{corollary:validity_algorithm1} and \ref{cor3}. The empirical coverage rates are very close to the intended confidence levels $p$. The bands for these parameters are not conservative.
 We know from Theorem \ref{theorem:bqte}
that the bands for the QE function are valid but may be conservative when the goal is to cover only the QE function independently
from the other functions. One of the objectives of the simulations is to assess if our QE-bands are narrow enough to be informative. The results in the sixth column of Table \ref{table:performance.count} show that the coverage rate of the bands is indeed larger than the theoretical coverage rate $p$  when the true QE function is uniformly $0$ (design 1) but is very close to $p$ when the distributions of the treated and control outcomes are different. The reason for this result is that the  Minkowski difference of two non-conservative confidence sets for two QF is \emph{not}   conservative for the difference in the parameters when (at least) one of the confidence set is a \emph{singleton}. While this case is irrelevant for continuous outcomes, it often happens for  discrete outcomes. As it    can be seen for instance in Figures \ref{figure:results_oregon_doc} or \ref{figure:results_testscores_9m}, the confidence bands for the QFs contains a single value at many probability indices.  Asymptotically, the bands for the QF of a discrete outcome will contain a single value at all quantiles except in the neighborhoods of the quantiles at which the QF jumps. Thus, asymptotically our bands for the QE function are not conservative except for the case when the QFs of $Y_1$ and $Y_0$ are identical, i.e. when $F^{\leftarrow}_1=F^{\leftarrow}_0$ uniformly. The second and third panels of the last column in Table \ref{table:performance.count} provide the empirical power of our bands to reject the null hypothesis that $F^{\leftarrow}_1=F^{\leftarrow}_0$. Even quite small deviations from the null hypothesis are detected  with relatively moderate sample sizes. As expected, the power increases with the sample size and with the deviation from the null hypothesis.  

Table \ref{table:performance.ordered} presents the results for an ordered outcome. $Y_0$ and $Y_1$ are both discretized random Gaussian variables that can take the values $\{0,1,...,5\}$. $Y_1$ is based on a latent standard Gaussian random variable while we consider three different latent variables for $Y_0$: $N(0,1)$, $N(0.2,1)$ and $N(0.4,1)$. The cut-off values are the same for both outcomes. They are chosen such that $Y_1$  takes the values $\{0,1,...,5\}$ with probability $\{0.1,0.16,0.24,0.24,0.16,0.1\}$ respectively. The results are extremely similar to the results in Table \ref{table:performance.count}: (i) the coverage rates for a single QF are very close to the intended coverage rate, (ii) the coverage rate for all QFs, DFs and the QE function is also very close to the intended rate, (iii) the coverage rate for the QE function is higher than the intended rate only when the true QE function is uniformly $0$, (iv) the power of our bands to reject an incorrect null hypothesis is substantial and increases in the sample size and the deviation from the null hypothesis. 

\newcolumntype{C}[1]{>{\centering\arraybackslash}m{#1}}
\begin{table}[ht]
\caption{Performance of the uniform bands for ordered outcomes}
\label{table:performance.ordered}
\centering
\begin{tabular}{r C{0.5cm} C{1.6cm} C{1.6cm} C{1.6cm} C{1.6cm} C{3cm}}
  \toprule
  \midrule
\multicolumn{1}{c}{$n$} & $p$ & \multicolumn{4}{c}{Empirical coverage probability for} & Prob. to reject\\
 & & $F_0$, $F^{\leftarrow}_0$ & $F_1$, $F^{\leftarrow}_1$ & all fct. & $F^{\leftarrow}_1-F^{\leftarrow}_0$  & $F^{\leftarrow}_1=F^{\leftarrow}_0$ \\
\hline
\multicolumn{7}{c}{Design 1: $Y^{*}_{0}\sim N(0,1)$ and $Y^{*}_{1}\sim N(0,1)$} \\
  400 & 0.99 & 0.98 & 0.98 & 0.98 & 1.00 & 0.00 \\ 
  400 & 0.95 & 0.94 & 0.94 & 0.93 & 1.00 & 0.00 \\ 
  400 & 0.90 & 0.89 & 0.88 & 0.88 & 1.00 & 0.00 \\ 
  1,600 & 0.99 & 0.98 & 0.99 & 0.99 & 1.00 & 0.00 \\ 
  1,600 & 0.95 & 0.94 & 0.95 & 0.94 & 1.00 & 0.00 \\ 
  1,600 & 0.90 & 0.89 & 0.90 & 0.89 & 1.00 & 0.00 \\ 
  6,400 & 0.99 & 0.99 & 0.99 & 0.99 & 1.00 & 0.00 \\ 
  6,400 & 0.95 & 0.95 & 0.95 & 0.95 & 1.00 & 0.00 \\ 
  6,400 & 0.90 & 0.89 & 0.90 & 0.90 & 1.00 & 0.00 \\ 
   \hline
\multicolumn{7}{c}{Design 2: $Y^{*}_{0}\sim N(0,1)$ and $Y^{*}_{1}\sim N(0.2,1)$}
\\
  400 & 0.99 & 0.98 & 0.98 & 0.98 & 0.99 & 0.02 \\ 
  400 & 0.95 & 0.94 & 0.94 & 0.93 & 0.95 & 0.11 \\ 
  400 & 0.90 & 0.89 & 0.88 & 0.87 & 0.90 & 0.21 \\ 
  1,600 & 0.99 & 0.98 & 0.98 & 0.98 & 0.98 & 0.62 \\ 
  1,600 & 0.95 & 0.94 & 0.95 & 0.94 & 0.94 & 0.88 \\ 
  1,600 & 0.90 & 0.89 & 0.89 & 0.89 & 0.89 & 0.95 \\ 
  6,400 & 0.99 & 0.99 & 0.99 & 0.99 & 0.99 & 1.00 \\ 
  6,400 & 0.95 & 0.95 & 0.95 & 0.95 & 0.95 & 1.00 \\ 
  6,400 & 0.90 & 0.89 & 0.90 & 0.90 & 0.90 & 1.00 \\ 
  \hline
\multicolumn{7}{c}{Design 3: $Y^{*}_{0}\sim N(0,1)$ and $Y^{*}_{1}\sim N(0.4,1)$} \\
  400 & 0.99 & 0.98 & 0.98 & 0.98 & 0.98 & 0.60 \\ 
  400 & 0.95 & 0.94 & 0.93 & 0.93 & 0.94 & 0.88 \\ 
  400 & 0.90 & 0.89 & 0.88 & 0.87 & 0.88 & 0.95 \\ 
  1,600 & 0.99 & 0.98 & 0.99 & 0.98 & 0.98 & 1.00 \\ 
  1,600 & 0.95 & 0.94 & 0.94 & 0.94 & 0.94 & 1.00 \\ 
  1,600 & 0.90 & 0.89 & 0.89 & 0.89 & 0.89 & 1.00 \\ 
  6,400 & 0.99 & 0.99 & 0.99 & 0.99 & 0.99 & 1.00 \\ 
  6,400 & 0.95 & 0.95 & 0.95 & 0.95 & 0.95 & 1.00 \\ 
  6,400 & 0.90 & 0.89 & 0.90 & 0.90 & 0.90 & 1.00 \\ 
  \midrule
   \bottomrule
   \multicolumn{7}{p{11cm}}{\scriptsize{\it Notes:} Based on $5,000$ simulations.}
\end{tabular} 
\end{table}

While our bands are--to the best of our knowledge--the only ones that have been proven to cover uniformly the QFs  and the QE functions of discrete outcomes, applied researchers may be tempted to use alternative heuristic approaches. For this reason, we compare the performance of our bands for the QE function with four alternative methods.\footnote{The results for the QFs are not shown because they are similar.}  We first experiment with directly bootstrapping the QE function and calculating sup-$t$ bands. However, the pointwise standard errors obtained via bootstrap are numerically equal to zero at many quantiles such that the $t$-statistic cannot be computed. We tried putting a lower bound on the pointwise standard errors to be able to calculate the $t$ statistics but this resulted in  extremely wide bands that always covered the true function. For this reason we do not report these results in the following tables. The second approach that we consider consists in bootstrapping the QE function and calculating constant width bands. This method avoids the need to divide by the estimated pointwise standard errors and could therefore be implemented.  The last two approaching are based on jittering (adding random noise) as suggested by \citet{machado2005quantiles} for count outcomes. We bootstrap the QE function of the smoothed outcomes and construct sup-$t$ bands  centered either around the smoothed QE function or around the original, unsmoothed QE function. \citet{machado2005quantiles}  show that standard methods can be used to make inference about the smoothed quantile function. On the contrary, we are interested in covering the original, unsmoothed QE function.

The results for the count outcomes are provided in Table \ref{table:comparison.count}, which compares the coverage probability of our new bands with that of the competing bands as well as the average length of the bands.\footnote{The computation time of these alternative methods is so high that we decided to not perform the simulations with $6,400$ observations. } The constant width bands obtained by bootstrapping directly the QE function are very conservative in all cases. Their average length is two to four times higher  than the average length of the bands that we have suggested. This bad behavior of the bootstrap for the QF of a discrete outcome comes at no surprise because it is known to be inconsistent for the estimation of the pointwise variance. \citet{huang1991estimating} finds in simulations that the bootstrap grossly overestimate the variance of the sample median of a discrete outcome, except when the QF jumps exactly at the median. The estimators based on jittering have the opposite problem: their coverage rate is below the intended rate and is even equal to zero for many distributions. The reason is simple: adding noise smoothes the differences over the whole range of quantiles such that the variance is underestimated where the QF jumps but is overestimated where the QF is flat. Note that these results do not contradict the results in \citet{machado2005quantiles}, which consider the smoothed QF as the true function, but show that adding noise to the outcome cannot help covering the unsmoothed QF.
Table \ref{table:comparison.ordered} presents the results of the simulations for the ordered outcomes. The conclusion are similar: bootstrapping the QE function directly leads to very wide bands while bootstrapping the jittered QE function leads to extreme undercoverage of the true function.

\begin{table}[ht]
\caption{Comparison with alternative bands for the QE fct.: count outcomes}
\label{table:comparison.count}
\centering
\begin{tabular}{r C{0.5cm} C{1cm} C{1cm} C{1cm} C{1cm} C{1cm} C{1cm} C{1cm} C{1cm}}
\toprule
\midrule
\multicolumn{1}{c}{$n$} & $p$ & \multicolumn{4}{c}{Coverage probability of the band:} & \multicolumn{4}{c}{Average length of the band:} \\
& & new & boot. & jitter1 & jitter2 & new & boot. & jitter1 & jitter2 \\
\hline
\multicolumn{10}{c}{Design 1: $Y_{0}\sim Poisson(3)$ and $Y_{1}\sim Poisson(3)$}
\\
  400 & 0.99 & 1.00 & 1.00 & 0.99 & 0.00 & 1.61 & 4.00 & 1.47 & 1.47 \\ 
  400 & 0.95 & 1.00 & 1.00 & 0.97 & 0.00 & 1.32 & 3.99 & 1.21 & 1.21 \\ 
  400 & 0.90 & 1.00 & 1.00 & 0.93 & 0.00 & 1.18 & 3.96 & 1.08 & 1.08 \\ 
  1,600 & 0.99 & 1.00 & 1.00 & 0.99 & 0.01 & 0.75 & 3.98 & 0.67 & 0.67 \\
  1,600 & 0.95 & 1.00 & 1.00 & 0.96 & 0.01 & 0.63 & 3.93 & 0.56 & 0.56 \\
  1,600 & 0.90 & 1.00 & 1.00 & 0.91 & 0.01 & 0.57 & 3.74 & 0.51 & 0.51 \\
\hline
\multicolumn{10}{c}{Design 2: $Y_{0}\sim Poisson(3)$ and $Y_{1}\sim Poisson(2.75)$}
\\
  400 & 0.99 & 1.00 & 1.00 & 0.01 & 0.00 & 1.56 & 3.99 & 1.44 & 1.44 \\ 
  400 & 0.95 & 0.98 & 1.00 & 0.00 & 0.00 & 1.28 & 3.81 & 1.19 & 1.19 \\ 
  400 & 0.90 & 0.94 & 1.00 & 0.00 & 0.00 & 1.15 & 3.47 & 1.06 & 1.06 \\ 
  1,600 & 0.99 & 0.99 & 1.00 & 0.00 & 0.00 & 0.73 & 2.76 & 0.66 & 0.66 \\
  1,600 & 0.95 & 0.96 & 1.00 & 0.00 & 0.00 & 0.61 & 2.26 & 0.55 & 0.55 \\
  1,600 & 0.90 & 0.91 & 1.00 & 0.00 & 0.00 & 0.55 & 2.08 & 0.50 & 0.50 \\
\hline
\multicolumn{10}{c}{Design 3: $Y_{0}\sim Poisson(3)$ and $Y_{1}\sim Poisson(2.5)$} \\
  400 & 0.99 & 0.99 & 1.00 & 0.02 & 0.00 & 1.52 & 3.92 & 1.42 & 1.42 \\ 
  400 & 0.95 & 0.97 & 1.00 & 0.00 & 0.00 & 1.26 & 3.32 & 1.17 & 1.17 \\ 
  400 & 0.90 & 0.93 & 1.00 & 0.00 & 0.00 & 1.13 & 2.75 & 1.05 & 1.05 \\ 
  1,600 & 0.99 & 0.99 & 1.00 & 0.00 & 0.00 & 0.72 & 2.34 & 0.65 & 0.65 \\
  1,600 & 0.95 & 0.96 & 1.00 & 0.00 & 0.00 & 0.61 & 2.07 & 0.54 & 0.54 \\
  1,600 & 0.90 & 0.92 & 1.00 & 0.00 & 0.00 & 0.55 & 2.02 & 0.49 & 0.49 \\
  \midrule
   \bottomrule
   \multicolumn{10}{p{11cm}}{\scriptsize{\it Notes:} Based on $5,000$ simulations.}
\end{tabular}
\end{table}

To summarize, for both types of discrete outcomes we come to the conclusion that the alternative methods either do not cover the true QE function with at least the chosen coverage rate or are much longer than the suggested bands. As an interesting by-product of these simulations, we note that  the average length of our bands converges to zero at the $\sqrt{n}$-rate.

\begin{table}[ht]
\centering
\caption{Comparison with alternative bands for QE: ordered outcomes} 
\label{table:comparison.ordered}
\begin{tabular}{r C{0.5cm} C{1cm} C{1cm} C{1cm} C{1cm} C{1cm} C{1cm} C{1cm} C{1cm}}
\toprule
\midrule
\multicolumn{1}{c}{$n$} & $p$ & \multicolumn{4}{c}{Coverage probability of the band:} & \multicolumn{4}{c}{Average length of the band:} \\
& & new & boot. & jitter1 & jitter2 & new & boot. & jitter1 & jitter2 \\
\hline
\multicolumn{7}{c}{Design 2: $Y^{*}_{0}\sim N(0,1)$ and $Y^{*}_{1}\sim N(0,1)$}
\\
  400 & 0.99 & 1.00 & 1.00 & 0.99 & 0.01 & 1.34 & 4.00 & 1.48 & 1.48 \\ 
  400 & 0.95 & 1.00 & 1.00 & 0.96 & 0.00 & 1.12 & 4.00 & 1.21 & 1.21 \\ 
  400 & 0.90 & 1.00 & 1.00 & 0.92 & 0.00 & 1.01 & 3.98 & 1.08 & 1.08 \\ 
  1,600 & 0.99 & 1.00 & 1.00 & 0.99 & 0.00 & 0.66 & 3.99 & 0.64 & 0.64 \\
  1,600 & 0.95 & 1.00 & 1.00 & 0.95 & 0.00 & 0.56 & 3.95 & 0.54 & 0.54 \\
  1,600 & 0.90 & 1.00 & 1.00 & 0.91 & 0.00 & 0.51 & 3.82 & 0.49 & 0.49 \\
\hline
\multicolumn{7}{c}{Design 2: $Y^{*}_{0}\sim N(0,1)$ and $Y^{*}_{1}\sim N(0.2,1)$}
\\
  400 & 0.99 & 0.99 & 1.00 & 0.07 & 0.01 & 1.33 & 3.94 & 1.50 & 1.50 \\ 
  400 & 0.95 & 0.95 & 1.00 & 0.01 & 0.00 & 1.11 & 3.57 & 1.22 & 1.22 \\ 
  400 & 0.90 & 0.90 & 1.00 & 0.00 & 0.00 & 1.01 & 3.17 & 1.09 & 1.09 \\ 
  1,600 & 0.99 & 0.98 & 1.00 & 0.00 & 0.00 & 0.66 & 2.34 & 0.65 & 0.65 \\
  1,600 & 0.95 & 0.94 & 1.00 & 0.00 & 0.00 & 0.56 & 2.06 & 0.55 & 0.55 \\
  1,600 & 0.90 & 0.89 & 1.00 & 0.00 & 0.00 & 0.50 & 2.01 & 0.50 & 0.50 \\
\hline
\multicolumn{7}{c}{Design 3: $Y^{*}_{0}\sim N(0,1)$ and $Y^{*}_{1}\sim N(0.4,1)$}
\\
  400 & 0.99 & 0.98 & 1.00 & 0.14 & 0.02 & 1.32 & 3.86 & 1.59 & 1.59 \\ 
  400 & 0.95 & 0.94 & 1.00 & 0.02 & 0.00 & 1.10 & 3.08 & 1.28 & 1.28 \\ 
  400 & 0.90 & 0.88 & 1.00 & 0.00 & 0.00 & 1.00 & 2.62 & 1.13 & 1.13 \\ 
  1,600 & 0.99 & 0.98 & 1.00 & 0.00 & 0.00 & 0.65 & 2.03 & 0.67 & 0.67 \\
  1,600 & 0.95 & 0.94 & 1.00 & 0.00 & 0.00 & 0.55 & 2.00 & 0.56 & 0.56 \\
  1,600 & 0.90 & 0.89 & 1.00 & 0.00 & 0.00 & 0.50 & 2.00 & 0.51 & 0.51 \\
  \midrule
   \bottomrule
   \multicolumn{10}{p{11cm}}{\scriptsize{\it Notes:} Based on $5,000$ simulations.}
\end{tabular}
\end{table}

\end{appendix}

\end{document}